\documentclass[cleanhead, cleanfoot,10pt,twocolumn]{config/asme2e}

\pdfoutput=1

\usepackage{empheq, amssymb}

\usepackage{lmodern}
\usepackage{amsmath}
\usepackage{amsfonts}
\usepackage{amssymb}

\usepackage{bm} 

\usepackage{txfonts}

\usepackage[T1]{fontenc}
\usepackage[utf8]{inputenc} 

\usepackage[dvipsnames]{xcolor} 

\usepackage[american]{babel} 

\usepackage{graphicx}

\usepackage{epstopdf}

\usepackage[noadjust]{cite} 

\usepackage{ltxcmds}

\usepackage{mathtools}

\usepackage{subcaption}

\usepackage{multirow} 

\usepackage{blindtext}

\usepackage{hyperref}

\usepackage{enumitem}

\usepackage{ifthen}

\usepackage{etoolbox}

\usepackage{cancel}

\usepackage{empheq}

\usepackage{fontawesome}

\usepackage{soul}

\usepackage{textcomp}

\usepackage{lipsum}

\definecolor{light-gray}{gray}{0.4}
\definecolor{box-gray}{gray}{1}

\hypersetup{
    unicode=false,          
    pdftoolbar=true,        
    pdfmenubar=true,        
    pdffitwindow=false,     
    pdfstartview={FitV},    
    pdftitle={Using High-fidelity Time-Domain Simulation Data to Construct Multi-fidelity State Derivative Function Surrogate Models for use in Control and Optimization},    
    pdfauthor={Athul K. Sundarrajan and Daniel R. Herber},     
    pdfsubject={},   
    pdfkeywords = {}, 
    pdfnewwindow=true,      
    colorlinks=true,
    allcolors=blue
}

\usepackage{nomencl}

\renewcommand\nomgroup[1]{%
  \item[\bfseries
  \ifstrequal{#1}{V}{ Variables}{%
  \ifstrequal{#1}{B}{ Subscripts}{%
  \ifstrequal{#1}{P}{ Notation}{%
  \ifstrequal{#1}{A}{ Acronyms}{}}}}]
}

\makenomenclature



\usepackage{dblfloatfix}
\usepackage{float}

\definecolor{block-gray}{gray}{0.95}

\usepackage[framemethod=TikZ]{mdframed}

\usepackage{xpatch}

\makeatletter
\xpatchcmd{\endmdframed}
  {\aftergroup\endmdf@trivlist\color@endgroup}
  {\endmdf@trivlist\color@endgroup\@doendpe}
  {}{}
\makeatother

\newcommand{\multifid}{\ensuremath{\hat{\bm{f}}_{\textrm{MF}}}}
\newcommand{\trad}{\ensuremath{\hat{\bm{f}}_{\textrm{NL}}}}
\newcommand{\actual}{\ensuremath{\bm{f}}}
\newcommand{\linfit}{\ensuremath{\hat{\bm{f}}_{\textrm{L}}}}
\newcommand{\taylor}{\ensuremath{\hat{\bm{f}}_{\textrm{T}}}}

\usepackage{fontawesome}

\usepackage{titlesec}

\usepackage{hyperref}

\newcommand{\doi}[1]{{doi:~\href{http://doi.org/#1}{#1}}\rmFullStop}

\newcommand*{\rmFullStop}{\rmifnextchar{.}{}{}}

\makeatletter
\newcommand{\rmifnextchar}[3]{%
  \begingroup
  \ltx@LocToksA{\endgroup#2}%
  \ltx@LocToksB{\endgroup#3}%
  \ltx@ifnextchar{#1}{%
    \def\next{\the\ltx@LocToksA}%
    \afterassignment\next
    \let\scratch= %
  }{%
    \the\ltx@LocToksB
  }%
}
\makeatother

\usepackage{colortbl}

\definecolor{light-gray}{gray}{0.6}

\newcommand{\xsection}[1]{\section[#1]{\MakeUppercase{#1}}}

\usepackage{algorithm2e}
\usepackage{xcolor}

\SetAlFnt{\footnotesize}

\SetAlCapSty{xAlCapSty}

\definecolor{commentcolor}{HTML}{1E4D2B}

\SetCommentSty{xCommentSty}

\SetNlSty{mynlfont}{}{} 

\LinesNumbered

\SetSideCommentRight

\DontPrintSemicolon

\RestyleAlgo{algoruled}

\usepackage{algorithm2e}
\usepackage{etoolbox}
\usepackage{xstring}
\usepackage{calc}
 
\newlength{\xalgowidth}
\newlength{\xalgoremainder}
\newlength{\xindentwidth}

\newenvironment{vAlgorithm*}[3][]{
  \setlength{\xalgowidth}{#2} 
  \setlength{\xindentwidth}{#3} 
  \setlength{\xalgoremainder}{\textwidth-\xalgowidth} 
  \SetCustomAlgoRuledWidth{\xalgowidth} 
  \IncMargin{\xindentwidth}
  \begin{algorithm*}[#1]
}
{
  \end{algorithm*} 
  \DecMargin{\xindentwidth}
}

{
  \end{algorithm} 
  \DecMargin{\xindentwidth}
}

\makeatletter
\patchcmd{\@algocf@start}{%
\begin{lrbox}{\algocf@algobox}%
}{%
\rule{0.5\xalgoremainder}{\z@}
\begin{lrbox}{\algocf@algobox}%
\begin{minipage}{\xalgowidth}%
}{}{}
\patchcmd{\@algocf@finish}{%
\end{lrbox}%
}{%
\end{minipage}%
\end{lrbox}%
}{}{}
\makeatother

\usepackage{accents}

\definecolor{needcolor}{HTML}{C62828}

\confshortname{IMECE2023}

\conffullname{Proceedings of the ASME 2023 \\ International Mechanical Engineering Congress and Exposition}

\confdate{November 2}
\confmonth{October 29 -}
\confyear{2023}
\confcity{New Orleans}
\confcountry{Louisiana}
\papernum{IMECE2023-112316}

\title{Using High-fidelity Time-Domain Simulation Data to Construct Multi-fidelity State Derivative Function Surrogate Models for use in Control and Optimization}

\author{Athul Krishna Sundarrajan\thanks{Corresponding author, \texttt{\href{mailto:athul.sundarrajan@colostate.edu}{athul.sundarrajan@colostate.edu}}}%
\affiliation{%
Graduate Student \\
Department of Systems Engineering \\
Colorado State University \\
Fort Collins, CO 80523 \\
\texttt{\href{mailto:athul.sundarrajan@colostate.edu}{athul.sundarrajan@colostate.edu}}}%
}

\author{
Daniel~R.~Herber%
\affiliation{%
Assistant Professor \\
Department of Systems Engineering \\
Colorado State University \\
Fort Collins, CO 80523 \\
\texttt{\href{mailto:daniel.herber@colostate.edu}{daniel.herber@colostate.edu}}}%
}

\usepackage{amsmath}

\begin{document}
 \setlength{\parskip}{0pt}
 \setlength{\parsep}{0pt}
 \setlength{\headsep}{0pt}

\setlength{\topsep}{0pt}

\abovedisplayshortskip=3pt
\belowdisplayshortskip=3pt
\abovedisplayskip=3pt
\belowdisplayskip=3pt

\titlespacing*{\section}{0pt}{18pt plus 1pt minus 1pt}{3pt plus 0.5pt minus 0.5pt}

\titlespacing*{\subsection}{0pt}{9pt plus 1pt minus 0.5pt}{1pt plus 0.5pt minus 0.5pt}

\titlespacing*{\subsubsection}{0pt}{9pt plus 1pt minus 0.5pt}{1pt plus 0.5pt minus 0.5pt}

\maketitle

\begin{abstract}\noindent%
\textit{Models that balance accuracy against computational costs are advantageous when designing dynamic systems with optimization studies, as several hundred predictive function evaluations might be necessary to identify the optimal solution.
The efficacy and use of derivative function surrogate models ({DFSMs}), or approximate models of the state derivative function, have been well-established in the literature.
However, previous studies have assumed an a priori state dynamic model is available that can be directly evaluated to construct the DFSM.
In this article, we propose an approach to extract the state derivative information from system simulations using piecewise polynomial approximations.
Once the required information is available, we propose a multi-fidelity DFSM approach as a predictive model for the system's dynamic response.
This multi-fidelity model consists of summation between a linear-fit lower-fidelity model and an additional nonlinear error corrective function that compensates for the error between the high-fidelity simulations and low-fidelity models.
We validate the model by comparing the simulation results from the DFSM to the high-fidelity tools.
The DFSM model is, on average, five times faster than the high-fidelity tools while capturing the key time domain and power spectral density~(PSD) trends.
Then, an optimal control study using the DFSM is conducted with outcomes showing that the DFSM approach can be used for complex systems like floating offshore wind turbines~(FOWTs) and help identify control trends and trade-offs.%
}%
\end{abstract}%

\vspace{1ex}

\xsection{Introduction}\label{sec:introduction}

Models that accurately capture the dynamics are needed to identify and understand system-level optimal designs. 
As a system's complexity and the fidelity of the underlying analyses increase, such models become more computationally expensive to evaluate~\cite{Deshmukh2017, Lefebvre2018}. 
For example, in highly-coupled multidisciplinary systems like floating offshore wind turbines (FOWT) and hydrokinetic turbines (HKT), the effect of the different subsystems (like the rotor, support structure (tower), floating platform, etc.) on each other must be fully considered to get an accurate response of the system~\cite{Pao2021}. 

To identify the optimal physical design and/or control law, evaluating the dynamic system several hundred or more times is necessary, which can make high-fidelity models impractical for some crucial use cases. 
Additionally, the software architecture of these system models might be such that it is impossible to link all the necessary variables of interest directly to an optimizer.
State variables are an example, as many simulation tools consider them internal, input-dependent quantities.
Engineers are also interested in understanding the impact of changing the system's physical parameters on its dynamics.
Tools like OpenFAST~\cite{openFAST} and WEC-Sim~\cite{Ruehl2022} can accurately capture the dynamics of renewable energy systems like wind turbines, hydrokinetic turbines, and wave energy converters.
However, for the various reasons discussed above, these tools have limitations regarding their use directly in optimization studies.
Therefore, computationally inexpensive system models that capture the dynamic response with sufficient accuracy to various input changes would be useful~\cite{AzadX1, Deshmukh2017, Sundarrajan2021}.

\subsection{Needs of Optimal Control and Control Co-design Studies}

One of these use cases is the design of control laws for dynamic systems through optimization studies, which have specific needs regarding the system models~\cite{Allison2014}.
Control design can be generally classified as either open loop or closed loop~\cite{Herber2014a}.
Open-loop optimal control studies aim to directly identify a control signal or trajectory that minimizes some goal(s) under various requirements. 
Closed-loop optimal control problems often aim to identify the controller gains or other tunable parameters within a particular feedback architecture.
Depending on the type of control study, different problem formulations, design variables, and solution strategies must be considered. 
For example, in robotics and autonomous vehicles, it is common to identify an optimal trajectory that satisfies all the constraints for the system while minimizing a cost function~\cite{Motallebiaraghi2023, Wang2019, Calzolari2017, Hoffmann2008}.
Closed-loop controllers are then developed to track this optimal trajectory. 
A prevalent practical numerical approach for both kinds of optimal control problems are direct methods such as the single-shooting (or simulation-based) and direct transcription (DT) approaches.
In these approaches, certain continuous signals are discretized on a finite number of points in a time grid~\cite{Herber2014a}.

The differences between the shooting and DT approaches primarily lie in what discretized signals are optimized and how the state dynamics are assured.
The shooting approach directly optimizes the discretized open-loop controls while assuring the system's dynamics through a sequential initial value problem solver, such as the family of Runge-Kutta methods.
Alternatively, the DT approach directly optimizes both the discretized open-loop controls and states as independent optimization variables while assuring the system's dynamics through constraints in the optimization problem \cite{Biegler2007}.
Therefore, the problem formulation, optimization variables, and method requirements differ between the two approaches.

Shooting-based approaches are easier to implement (i.e., can utilize basic input/output dynamic modeling paradigms) with repeat simulations of the dynamics for different values explored by the optimizer and are widely used for the closed-loop control design of dynamic systems.
However, shooting methods can have convergence issues and struggle to effectively handle additional path constraints when they are added to the problem~\cite{Biegler2007}.
On the other hand, DT methods can handle path constraints efficiently, among other benefits, but generally require an internal state dynamic model (see Sec.~\ref{sec:dfsm-prelim})~\cite{Allison2014, Sundarrajan2021a}.
Because open-loop approaches do not assume a fixed control architecture, the optimizer has more degrees of freedom to identify the optimal dynamic response of the system-of-interest~\cite{Nash2020, Nash2021, Deshmukh2015a}.
This principle can be used to identify system-level optimal designs for dynamic systems~\cite{Deshmukh2015a}.

Furthermore, in many dynamic systems, there is significant coupling between the plant parameters and the control inputs~\cite{Allison2014, Azad2019}. 
To identify system-level optimal designs, this plant-controller coupling must be taken into account~\cite{Fathy2003a}.
The concurrent optimization of the plant and controller is also known as control co-design (CCD)~\cite{Allison2014, GarciaSanz2019a,Herber2019a}. 
For emerging renewable energy technologies, experts have recognized the potential of CCD to identify designs with a lower cost of energy that can be adopted by the industry~\cite{GarciaSanz2019a, Jonkman2021, Ross2022}.
However, CCD approaches require practical, structured, and efficient dynamic system models that can predict changes to both the control and plant aspects.


The focus of this article is on developing computational modelsfrom high-fidelity simulations that balance cost and accuracy and can be used to solve optimal control problems, with a particular emphasis on the needs of open-loop optimal control as there has been less attention towards approximate models for these problems.
The model must be constructed in such a form that both shooting and direct transcription approaches can be used efficiently. 
Although the focus is on open-loop methods, the modeling approach presented can also be applied in the context of closed-loop optimal control problems.

The rest of the article is organized as follows.
In Sec.~\ref{sec:background}, we provide a brief overview of different surrogate modeling approaches, including derivative function surrogate modeling~(DFSM), which is the main focus of this article.
In Sec.~\ref{sec:DFSM}, we discuss the assumptions and the steps involved with the multi-fidelity DFSM approach through an illustrative example and validate the approach by comparing the predictions made by the DFSM to the actual derivative function.
In Sec.~\ref{sec:FOWT-application}, we use the DFSM model to construct and solve open-loop optimal control problems for a FOWT.
In Sec.~\ref{sec:conclusion}, we summarize the approach and results from this article and provide directions for future work.

\xsection{Background}\label{sec:background}
\subsection{Surrogate Models}
\label{sec:low-fidelity-model}

An often-used approach for constructing an approximate model of a real (or potential) engineered object is based on physics first principles, such as linear dynamic behavior of key energy efforts and flows.
Then parameters in the linear relationships (e.g., the stiffness constant of force-displacement relationship) can be found using analytic expressions or data and system response outputs predicted for different inputs.
These simplified relationships can yield to more accurate ones that better represent the behavior of the system of interest.
This additional model accuracy can lead to detailed mathematical representations closely approximating reality but may be computationally expensive to evaluate.
Determining in a computationally-effective way how outputs change for a suitably accurate model form (e.g., transfer function, linear state space model, or nonlinear system) is valuable in engineering dynamic systems.

Surrogate models have found use in approximating computationally-expensive functions, and their application in optimization-based design studies is well studied in the literature~\cite{Forrester2009}.
Computationally-inexpensive metamodel forms can be trained to capture the input-output response of expensive functions~\cite{SMT2019}. 
Then, these trained metamodels can serve as the objective and constraint functions in an optimization study, with the inputs being the optimization variables under consideration.
There are various ways of constructing surrogate models, with some now being discussed.

\subsubsection{Generalized Linear Metamodels.\label{sec:linear-metamodels}}~Some of the simplest surrogate model forms are linear.
Coefficients in a linear relationship might be found using linear regression techniques.
For dynamic systems, linear state-space models and transfer functions can be constructed using time-domain or frequency-domain data \cite{Overschee1996,Kollar1997}.
These approaches often fall under system identification methods.
Given input-output data and a set of candidate models, these approaches seek to identify the model and its associated parameters that minimize the error between the actual output and predicted output.

A popular alternative (including in control-design studies) is based on using first-order Taylor-series approximations~\cite{Hendricksa}.
More specific linear forms, such as the linear-parameter varying (LPV) models discussed in Ref.~\cite{Sundarrajan2021}, are possible as well. 
However, computing the Taylor-series expansion can be computationally expensive and sensitive.
Furthermore, modeling the changes in plant variables through these linear models can be challenging and local, thereby limiting their use in CCD studies~\cite{Jonkman2022}.

\subsubsection{Generalized Nonlinear Metamodels.\label{sec:deep-learning}}



The limitations of linear forms has led to significant work on nonlinear surrogate models.
Nonlinear surrogate modeling and system identification methods using time-series data for control and simulation are well-established~\cite{Worden2018, Kerschen2006, Deese2020}, including software packages~\cite{Ayala2020}.
These might utilize nonlinear metamodels forms like radial basis functions~(RBFs), neural networks~(NNs), and Gaussian process regression~(GPR). 

Recently, the predictive capabilities of deep neural networks (DNNs), especially recurrent neural networks (RNNs), long short-term memory (LSTM) networks, continuous-time echo state networks (CTESNs), and autoencoders, have been used to approximate nonlinear dynamic system response~\cite{Roberts2022, anantharaman2020, Lee2021, Zhao2022}.
These approaches have also been referred to as neural simulations, where a DNN is trained to predict the dynamic response of a nonlinear dynamic system.
In addition to these advances, several authors have used deep-learning paradigms to construct neural state-space models~\cite{Gedon2021, Chakrabarty2022}.
For example, Ref.~\cite{Chen2018OptimalCV} uses neural state-space models to perform model predictive control~(MPC) for building an HVAC system.
The efficacy of these approximate models in various CCD tasks still needs further exploration as most of these approaches do not consider plant changes as a part of the data-driven model, and the time horizons are often quite long in open-loop optimal control CCD studies.

\subsubsection{Multi-fidelity~Approaches.}~Multi-fidelity approaches use different combinations of models to approximate a computationally-expensive high-fidelity response.
Grey-box models, used extensively for the control design of building energy systems, can be considered a specific type of multi-fidelity model~\cite{Sohlberg2008}.
Multi-fidelity models are used extensively for the design of complex engineering applications like airfoils, wind turbine blades, layout optimization of wind turbine farms, and topology optimization~\cite{March2012, Jasa2022}.
Please refer to Refs.~\cite{Peherstorfer2018, March2012} for a detailed summary of multi-fidelity modeling and Refs.~\cite{Sohlberg2008, Li2021, Bacher2011} for grey-box models and their various applications.

One basic form of a multi-fidelity model is as follows:
\begin{align}
    \label{eq:multi-fidelity}
    f_{\textrm{high}}(\bm{x}) \approx \hat{f}_{\textrm{low}}(\bm{x}) + \hat{f}_{\textrm{med}}(\bm{x})
\end{align}

\noindent where the low-fidelity model $\hat{f}_{\textrm{low}}$ and medium-fidelity model $\hat{f}_{\textrm{med}}(\bm{x})$ are combined to approximate the high-fidelity model $f_{\textrm{high}}$.
There are different ways to aggregate the low/mid-fidelity models, but we consider them additive in this article.

The form in Eq.~(\ref{eq:multi-fidelity}) implies that the exclusion of the $\hat{f}_{\textrm{med}}$ results in a lower, but still useful, model of $f_{\textrm{high}}$.
For example, one common multi-fidelity approach (and the one considered in this article) is to utilize a linear metamodel for $\hat{f}_{\textrm{low}}$ and a nonlinear metamodel for $\hat{f}_{\textrm{med}}$.
Then, if it is found that $\hat{f}_{\textrm{med}}$ is negligible for some relationship, then the identified exclusive linear mapping can help reduce the time taken to construct the model as well as solution time, as discussed in Ref.~\cite{Herber2020d} for nonlinear open-loop optimal control problems.


\subsection{DFSM Preliminaries\label{sec:dfsm-prelim}}



For dynamic systems, an ordinary differential equation~(ODE) describes how the states evolve given the states, inputs, and parameters. 
In this article, this function will be referred to as the state derivative function or simply the derivative function.
Furthermore, various important outputs might also be modeled. 
Now, consider a nonlinear state-space model of the following form:
\begin{subequations} \label{eq:state-space}
\begin{align}
    \label{eq:deriv-fun}
    \frac{d\bm{\xi}}{dt} = \dot{\bm{\xi}} = \bm{f}(\bm{\xi}(t), \bm{u}(t), \bm{p}) \\
    \bm{y}(t) = \bm{g}(\bm{\xi}(t), \bm{u}(t), \bm{p}) \label{eq:outputs}
\end{align}
\end{subequations}

\noindent where $t$ is time, $\bm{\xi}(t)$ are the states under consideration, $\bm{u}(t)$ are the controls, $\bm{p}$ are the static parameters, and $\bm{y}(t)$ are the outputs\footnote{The time dependence of the input/output variables will not always be explicitly shown for conciseness.\vspace{1\baselineskip}}.
The derivative function is then $\bm{f}(\cdot)$ and output function $\bm{g}(\cdot)$.

In certain cases, evaluating $\bm{f}$ is the most computationally-expensive operation in the study (when it is directly available).
Constructing a surrogate model for this function can reduce this computational expense while still capturing the dynamic state evolution.
Such an approach has been studied under the term \textit{derivative function surrogate model}~(DFSM)~\cite{Deshmukh2017, Lefebvre2018, Zhang2022, Qiao2021}.
The goal of any DFSM approach is to construct a surrogate model $\hat{\bm{f}}$ of the function $\bm{f}$ to predict the state-derivative values $\dot{\bm{\xi}}$, given the inputs $\bm{I} = \left[\bm{u},\bm{\xi}\right]$:
\begin{align}
    \label{eq:dfsm}
    \dot{\bm{\xi}} = \bm{f}(\bm{I}) \approx \hat{\bm{f}}(\bm{I})
\end{align}

\noindent If a model of the form as shown in Eq.~(\ref{eq:deriv-fun}) is available, $\bm{f}(\cdot)$ is sampled to obtain input-output pairs $\langle\bm{I},\dot{\bm{\xi}}\rangle$, and data-fitting methods like RBF and GPR are used to construct $\hat{\bm{f}}$.
We use the notation \trad~to denote traditional DFSM approaches as presented in Refs.~\cite{Deshmukh2017, Lefebvre2018, Zhang2022, Qiao2021}.
Once the DFSM is constructed, it can be readily used in studies that require $\dot{\bm{\xi}}$ to be computed, such as DT-based and shooting-based optimal control studies.
The approaches presented in the literature assume that the derivative function (and therefore $\dot{\bm{\xi}}$) is available such that $\bm{f}$ can be directly queried for any set of inputs.
This assumption might not hold for some system models and simulation tools, but alternate strategies will be developed here.

\subsection{WEIS Toolbox for Modeling Floating Offshore Wind Turbines (FOWTs)}
\label{sec:WEIS-description}

The motivating application for the methods in this article centers around tools for the development of state-of-the-art wind turbines.
In particular, WEIS is an open-source tool developed primarily by the National Renewable Energy Laboratory~(NREL) to enable CCD of wind turbines~\cite{Jonkman2021}.
It is built on OpenFAST~\cite{openFAST}, an aero-servo-hydro dynamic solver that can be used to evaluate the dynamic response of wind turbines.
WEIS can perform CCD studies at three different levels of fidelity.
At level 1, it uses a frequency-domain model to approximate the dynamic response.
At level 2, WEIS uses a linearized time-domain model; at level 3, a user has full nonlinear time-domain simulations available using OpenFAST.

The inputs WEIS requires to run the simulations are the degrees of freedom considered in the given model, the description of the wind input, and the fidelity level of the dynamic model.
Different design load cases (DLCs) have been specified by the International Electrotechnical Commission~(IEA) to test wind turbines under different scenarios.
Simulating a single DLC takes an average of $15$ minutes.
WEIS is a fitting use case for DFSM, as the user does not have direct access to the underlying dynamic model, and direct simulations are generally considered too expensive for comprehensive optimal control and CCD studies.



\xsection{Multi-fidelity DFSM}\label{sec:DFSM}

In this section, we describe the assumptions with the DFSM approach, and the different steps involved in constructing the DFSM.

\subsection{Overview}\label{sec:overview}

Here we assume that the derivative function in Eq.~(\ref{eq:deriv-fun}) is not available in a form that can be evaluated directly, but a black box code can be simulated for a given input $\bm{u}\in\mathbb{R}^{n_u}$.
We assume that the states $\bm{\xi}\in\mathbb{R}^{n_{\xi}}$ are available from the outputs of the simulation $\bm{y}\in\mathbb{R}^{n_y}$, and the model does not have any other internal states, such that $\bm{\xi} \subset \bm{y}$.
To approximate $\bm{f}(\cdot)$ we consider a multi-fidelity model from Eq.~(\ref{eq:multi-fidelity}) of the form:
\begin{align}
    \bm{f}(\cdot) \approx \hat{\bm{f}}_{\textrm{low}}(\cdot) + \bm{e}(\cdot)
\end{align}

\noindent
where $\hat{\bm{f}}_{\textrm{low}}(\cdot)$ is a low-fidelity linear-fit model, and $\bm{e}(\cdot)$ is a higher-fidelity component that attempts to approximate the remaining error between $\bm{f}(\cdot)$ and $\hat{\bm{f}}_{\textrm{low}}(\cdot)$.
The steps in the multi-fidelity DFSM approach are outlined first:%
\begin{enumerate}

\item Run the necessary simulations to obtain the baseline data for state and output trajectories (see Sec.~\ref{sec:generate-simulations}).

\item Construct at least a $C^1$ continuous polynomial approximation of the state trajectories $\hat{\bm{\xi}}(t)$ and then evaluate polynomial approximation derivative $\hat{\dot{\bm{\xi}}}(t)$ (see Sec.~\ref{sec:poly-approx}).

\item Using the input-output data, construct a least-squares linear-fit approximation creating $\hat{\bm{f}}_{\textrm{low}}$ (see Sec.~\ref{sec:construct-linear-fit}).

\item Using the input-output data, evaluate the remaining error between the actual state derivatives and the linear-fit model (see Sec.~\ref{sec:k-means-sampling}).

\item Train a nonlinear surrogate model on this error using a selected approach determining $\bm{e}$ (see Sec.~\ref{sec:construct-corrective-function}).

\item Validate the resulting multi-fidelity model (see Sec.~\ref{sec:validation}).

\end{enumerate}

\subsubsection{Illustrative Example.}~Consider the derivative function of the two-link robot system described in Example 2.10 of Ref.~\cite{Hendricksa}.
The system is characterized by the angle of the two links~($[\theta_1,\theta_2]$), their inertial velocities~($[\dot{\theta}_1,\dot{\theta}_2]$), the torques applied at the links~($[u_1,u_2]$), and the physical characteristics of the system like the length, mass, and moment of inertia of the two links represented as $\bm{p}$.
Assuming the states to be $\bm{\xi} = [\theta_1,\dot{\theta}_1,\theta_2,\dot{\theta}_2]^T$, and controls to be $\bm{u} = [u_1,u_2]^T$, the nonlinear state-space model be represented as:
\begin{subequations}
\begin{align}
\label{eq:two-link-robot}
\dot{\bm{\xi}} = \bm{f} &= \begin{bmatrix}
\xi_2 \\
f_2(\bm{\xi},\bm{u},\bm{p}) \\
\xi_4 \\
f_4(\bm{\xi},\bm{u},\bm{p})
\end{bmatrix}\\
\bm{y} &= \bm{\xi}
\end{align}
\end{subequations}

\noindent The expressions for $f_2$ and $f_4$ are highly nonlinear and derived from the Lagrangian of the robotic system.
For example, using fixed $\bm{p}$, $f_2$ is:
\begin{align}
\label{eq:f2_full}
f_2(\bm{\xi},\bm{u}) &= -\Big(5500(u_1 - u_2) - 100062\cos(\xi_1) + \cdots \notag \\
& + 29430\cos(\xi_1 + 2\xi_3) + 1800\xi_2^2\sin(2\xi_3) + \cdots \notag \\ & - 6000u_2\cos(\xi_3) + 3300\xi_2^2\sin(\xi_3) + 3300\xi_4^2\sin(\xi_3)  + \cdots \notag \\ 
& + 6600\xi_2\xi_4\sin(\xi_3)\Big)/(1800\cos(2\xi_3) - 5295)
\end{align}
Please refer to Ref.~\cite{Hendricksa} for the detailed derivation of $\bm{f}$ and values of $\bm{p}$.
This analytic system will be used to demonstrate the different steps associated with the multi-fidelity DFSM approach.

\subsection{Generating Simulations}
\label{sec:generate-simulations}

For the system considered in Eq.~(\ref{eq:two-link-robot}), we generate total of $n_{\textrm{sim}}$ starting points $\bm{\xi}(0)$ and control inputs $\bm{u}$.
In this example, we use uniform random signals for $\bm{u}$, but a more strategic selection of controls might be considered that explore the state space. 
By simulating the system using these inputs, we can obtain $n_{\textrm{sim}}$ sets of output trajectories $\bm{y}(t)$ and time mesh $\bm{t}$.
From $\bm{y}$, the state trajectories $\bm{\xi}$ can be extracted and organized as:
\begin{subequations}
\begin{align}
    \bm{T} &= \begin{bmatrix} \bm{t}^{(1)} & \bm{t}^{(2)} & \cdots & \bm{t}^{(n_{\textrm{sim}})}\end{bmatrix}\\
    \bm{I} = \begin{bmatrix} \bm{U} \\ \bm{X} \end{bmatrix} &=  \begin{bmatrix}\bm{u}^{(1)} & \bm{u}^{(2)} & \cdots & \bm{u}^{(n_{\textrm{sim}})}  \\  \bm{\xi}^{(1)} & \bm{\xi}^{(2)} & \cdots & \bm{\xi}^{(n_{\textrm{sim}})} \end{bmatrix} 
\end{align}
\end{subequations}

\subsection{Extracting State Derivative Information}
\label{sec:poly-approx}

When the direct evaluation of $\bm{f}$ is not possible, the state derivative information can be indirectly obtained from the simulated state trajectories.
A continuous polynomial approximation of sampled signals is available in many tools.
With at least a $C^1$ approximation, approximate first-order derivatives can be obtained (and higher-order derivatives as well, depending on the method used).
Here, a cubic spline polynomial approximation is used to construct continuous $\bm{\xi}(t)$ and then the exact polynomial derivatives are found for  $\dot{\bm{\xi}}(t)$.
Once $\bm{\xi}$ and $\bm{t}$ are available, the functions \texttt{spline} and \texttt{fnder}, available in the curve fitting toolbox in \texttt{MATLAB}, and the class \texttt{CubicSpline} from \texttt{SciPy} can be used to construct the approximation and evaluate the derivatives.

\begin{figure}[t]
\centering
\includegraphics[width = 0.4\textwidth]{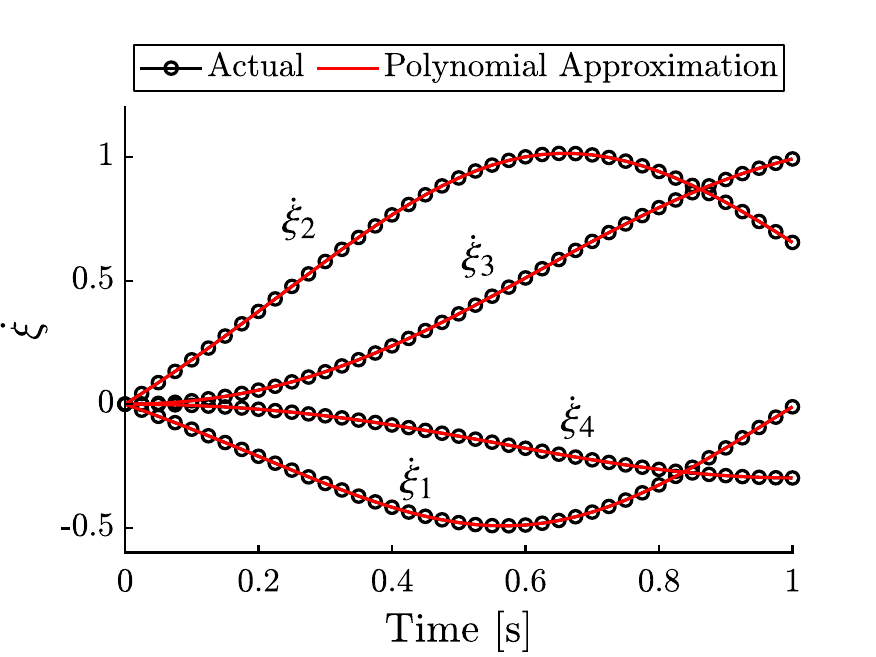}
\caption{Comparison of actual derivative value to the polynomial approximation for the two-link robot system.}
\label{fig:dx-comparison} %
\end{figure}

For the illustrative example, Fig.~\ref{fig:dx-comparison} compares the actual state derivatives for Eq.~(\ref{eq:two-link-robot}) and the ones obtained from evaluating the polynomial approximation.
The mean error between the actual derivatives and the polynomial derivative approximation evaluated for a hundred random simulations is of the order $10^{-7}$, which shows that the polynomial approximation can provide accurate derivative values on a simple analytic example.

Now, the input-output pairs $\langle\bm{I},\dot{\bm{X}}\rangle$ are available to construct $\hat{\bm{f}}(\cdot)$ where:
\begin{align}
    \dot{\bm{X}} = \begin{bmatrix} \dot{\bm{\xi}}^{(1)} & \dot{\bm{\xi}}^{(2)} & \cdots & \dot{\bm{\xi}}^{(n_{\textrm{sim}})} \end{bmatrix}
\end{align}

\subsection{Low-Fidelity Model}
\label{sec:construct-linear-fit}

The low-fidelity portion is found by constructing a least-squares approximation between the inputs $\bm{I}$ and the state derivatives $\dot{\bm{X}}$:
\begin{subequations}
   \begin{align}
    \label{eq:multi-fidelity-dfsm}
    \hat{\bm{f}}_{\textrm{low}}(\bm{I}) = \hat{\bm{f}}_{\textrm{L}}(\bm{I}) &= \bm{L}\bm{I} \\
    \bm{L} = \bm{I}^T (\bm{I} \bm{I}^T)^{-1} \dot{\bm{X}}
\end{align} 
\end{subequations}

\noindent The corresponding state $\bm{A}_{\textrm{L}}$ and input $\bm{B}_{\textrm{L}}$ matrices can be extracted from $\bm{L}$.
This model will be referred to as a `linear-fit' model as opposed to a `linear' or `linearized' to avoid ambiguity.

In addition to the inputs $\bm{I}$, some systems are characterized by additional parameters $\bm{w}$ that affect the dynamics.
Wind turbine dynamics, for example, depend heavily on the incoming wind speed.
A single linear-fit model might not be adequate to capture the dynamics of wind turbines over the entire range of wind speed values.
In such cases, multiple linear-fit models can be derived for a range of parameter values and aggregated to provide a more accurate prediction, such that the \linfit{} is a function of $\bm{w}$.
This is similar to the idea of linear-parameter varying (LPV) models explored in detail in Ref.~\cite{Sundarrajan2021a}, where the authors have shown that utilizing multiple linear models can provide a better approximation as opposed to using a single linear model with:
\begin{align}
    \label{eq:LPV-formulation}
    \hat{\bm{f}}_{\textrm{L}} = \bm{L}(\bm{w})\bm{I} = \begin{bmatrix}\bm{B}_{\textrm{L}}(\bm{w})& \bm{A}_{\textrm{L}}(\bm{w})\end{bmatrix}\bm{I}
\end{align}

\subsection{Subsampling for DFSM Construction}
\label{sec:k-means-sampling}

With $\hat{\bm{f}}_{\textrm{L}}$ determined the remaining error is calculated as:
\begin{align}
    \bm{E} = \dot{\bm{X}} - \bm{L}\bm{I}
\end{align}


\noindent In this article, we select an $\bm{e}$ that, unlike the least-squares linear-fit operation, does not scale as well when the number of data points and input dimension increases (see Sec.~\ref{sec:construct-corrective-function}).
Therefore, there is a need to strategically select some of the data from $\langle\bm{I},\bm{E} \rangle$.

Previous DFSM studies have directly generated input samples to evaluate the high-fidelity model using space-filling approaches like Latin hypercube sampling \cite{Deshmukh2017,Qiao2021}.
In this study, this approach would not be feasible based on some of the initial assumptions that $\bm{f}$ is not directly available.
Then, the goal is to extract samples from the simulated data that cover the region of interest.
Some studies mentioned in Sec.~\ref{sec:deep-learning} that have worked with simulated data addressed this problem using the `max-min algorithm' from Ref.~\cite{Johnson1990}.
The max-min approach aims to select a set of points from the simulated data covering the entire domain.
The approach is to maximize the minimum distance between the samples iteratively, starting from an initial random set.

In this study, $k$-means clustering is used as the approach for subsampling the given data with the $k$-means++ algorithm for selecting the initial value of the clusters~\cite{Arthur2007kmeansTA}.
Similar to the max-min algorithm, the goal is to find $n_s$ representative clusters in the given data such that the distance between the points in a given cluster is minimized and the distance between the centroids of the clusters is maximized.
The centroid values are used as representatives for each cluster with their associated $\dot{\bm{X}}_i$ being the arithmetic mean of all data points in the identified cluster $i$.

\subsection{Constructing the Nonlinear Error Corrective Function}
\label{sec:construct-corrective-function}

Radial basis functions (RBFs) are used to construct the nonlinear error corrective function $\bm{e}$ in this study, although many approaches may be considered.
Using RBFs to approximate nonlinear functions is a well-established way of constructing surrogate models.
Their use in different DFSM and multi-fidelity modeling studies has also been well established~\cite{March2012, Deshmukh2017}.
Briefly, RBF models approximate a function $F$ as:
\begin{align}
    \label{eq:rbf}
    F(\bm{x}) = \sum_{i = 1}^N w_i \cdot \phi(\lVert \bm{x} - \bm{x}_i \rVert_2)
\end{align}

\noindent where ${\phi}(\cdot)$ is the underlying basis function, and $\bm{w}$ is the array of weights that can be optimized for increased accuracy.
In this study, we consider a Gaussian basis function such that $\phi(\bm{x}) = \textrm{exp}(-\bm{x}^2)$.
Unlike other approaches, RBFs have the advantage of being robust and have fewer parameters that need to be tuned.
We use RBF as part of an RBF network, a specific type of neural network with radial basis functions as neurons.
The function \texttt{newrb} in the deep learning toolbox in \texttt{MATLAB} is used for this step~\cite{RBF}.
With $\bm{e}$ available, we have the multi-fidelity DFSM model from Eq.~(\ref{eq:multi-fidelity-dfsm}) that approximates \actual.

\subsection{Creating Surrogates of Outputs}
\label{sec:outputs-sm}

For many systems, additional outputs are also important.
Similar to the derivative function, surrogate models can be constructed for the outputs $\bm{y}$.
For each simulation in $n_{\textrm{sim}}$, the corresponding outputs can be obtained as:
\begin{align}
    \bm{Y} = \begin{bmatrix}\bm{y}^{(1)} & \bm{y}^{(2)} & \cdots & \bm{y}^{(n_{\textrm{sim}})} \end{bmatrix}
\end{align}

Then, the steps previously outlined in Secs.~\ref{sec:construct-linear-fit} to \ref{sec:construct-corrective-function} can be used to construct a surrogate model of the output function $\bm{g}$ in Eq.~(\ref{eq:outputs}) such that:
\begin{subequations}
    \begin{align}
   \bm{y} =  \bm{g}(\bm{I})  \approx \hat{\bm{g}}(\bm{I}) \\
   \hat{\bm{g}} = \begin{bmatrix}
       \bm{D}_{\textrm{L}}&\bm{C}_{\textrm{L}}
   \end{bmatrix}\bm{I} + \bm{e}_g(\bm{I})
\end{align}
\end{subequations}

\noindent with  $\bm{C}_{\textrm{L}}$ and $\bm{D}_{\textrm{L}}$ denoting the corresponding linear-fit model matrices for the outputs.
With $\hat{\bm{g}}$, we have a surrogate model for the entire state-space model of a given system outlined in Eq.~\ref{eq:state-space}.

\subsection{DFSM Validation\label{sec:validation}}

Previous DFSM studies made limited attempts to validate the DFSM model before using it within optimal control studies.
This is because adaptive refinement was used where the DFSM is successively updated to be more accurate around the trajectories identified by the optimal control study by adding more points.
A drawback of this strategy is that the DFSM is sensitive to the initial sampling and could be overfitted around the optimal trajectory.

Here, we propose using time-domain simulations to validate the predictive capability of DFSM by comparing the simulation results with the actual model response over a wide range of inputs.
To illustrate the validation approach, we again use the two-link robot example described by Eq.~(\ref{eq:two-link-robot}) to validate the DFSM.

\begin{figure*}[ht]
\centering
\includegraphics[height=0.22in]{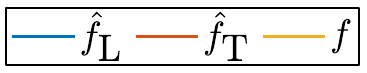}\\
\begin{subfigure}[t]{0.25\textwidth}
\centering
\noindent\includegraphics[scale=0.34]{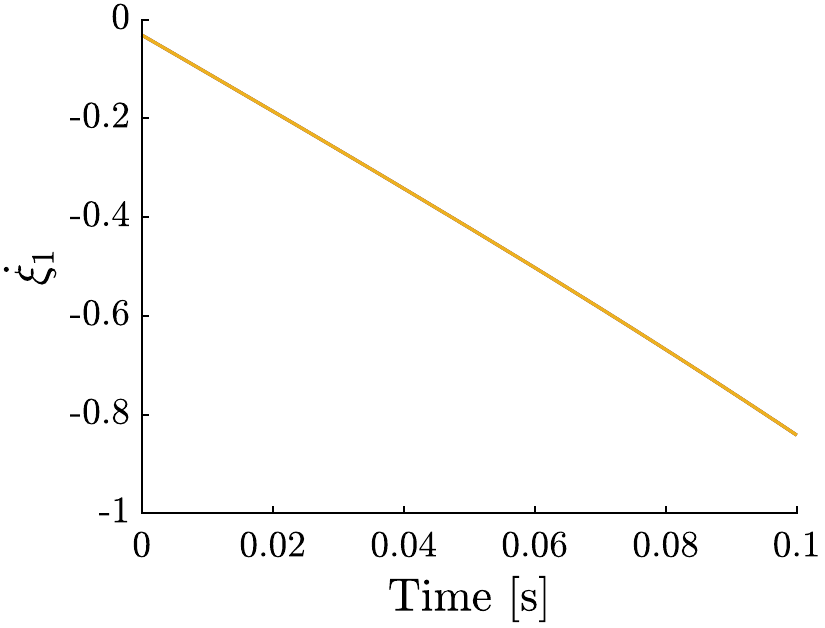}
\caption{$\dot{\xi}_1(t)$.}
\label{fig:dx_lin_1} 
\end{subfigure}%
\begin{subfigure}[t]{0.25\textwidth}
\centering
\noindent\includegraphics[scale=0.34]{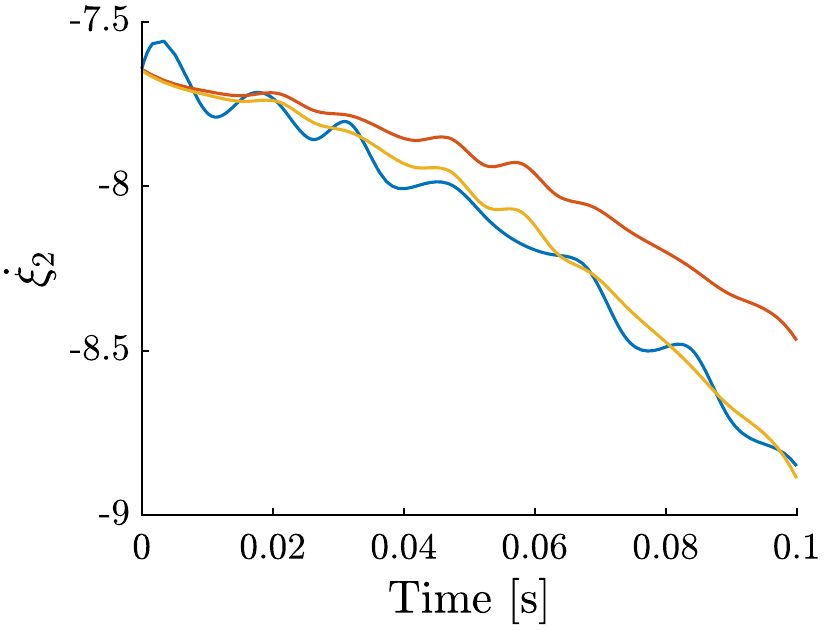}
\caption{$\dot{\xi}_2(t)$.}
\label{fig:dx_lin_2} 
\end{subfigure}%
\begin{subfigure}[t]{0.25\textwidth}
\centering
\noindent\includegraphics[scale=0.34]{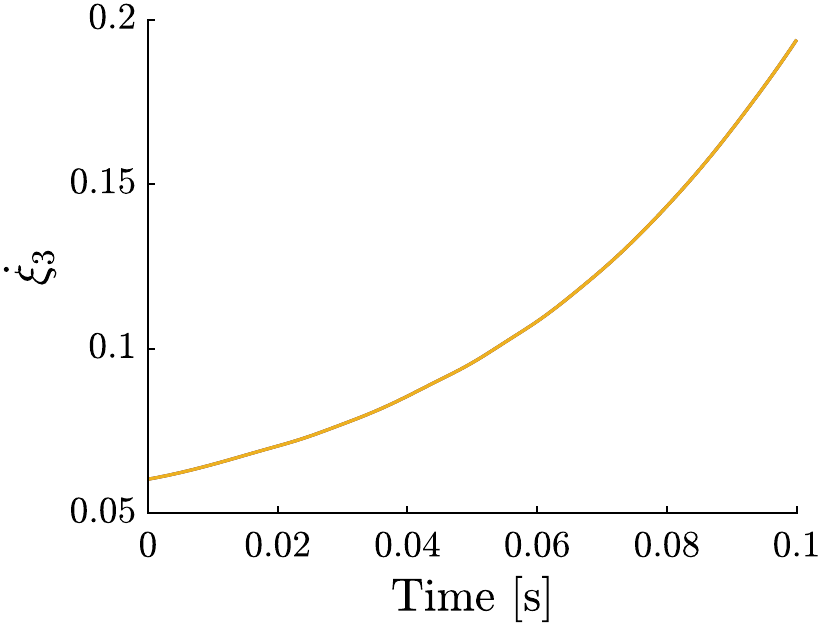}
\caption{$\dot{\xi}_3(t)$.}
\label{fig:dx_lin_3} 
\end{subfigure}%
\begin{subfigure}[t]{0.25\textwidth}
\centering
\noindent\includegraphics[scale=0.34]{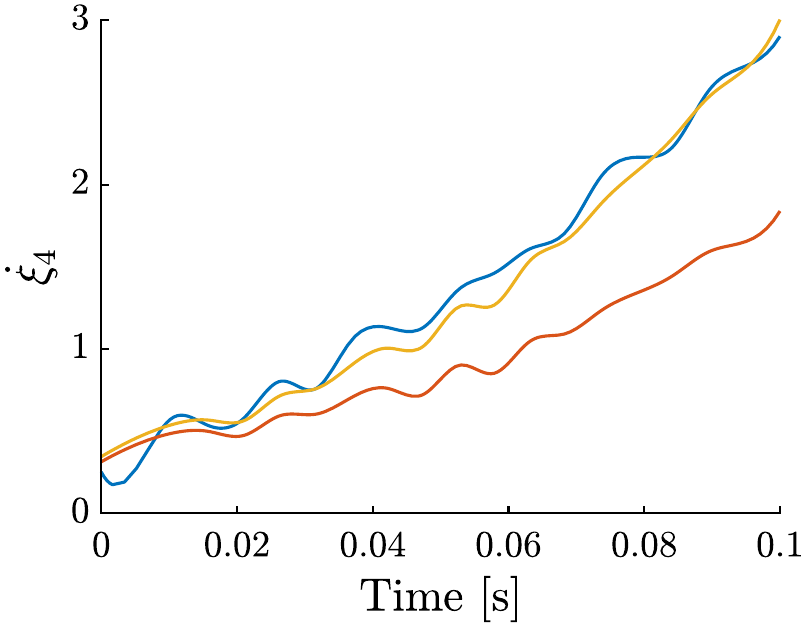}
\caption{$\dot{\xi}_4(t)$.}
\label{fig:dx_lin_4} 
\end{subfigure}%
\caption{Comparison of the state derivatives predicted by  \linfit~and \taylor~to \actual~for a test simulation.}
\label{fig:dx_lin_comp}
\end{figure*}

\subsubsection{Validating the Linear-fit Model.\label{sec:linfit validation}}~We start by comparing the linear-fit model to the first-order Taylor series approximation of Eq.~(\ref{eq:two-link-robot}) around the stationary point $\bm{\xi}_{\textrm{stat}} = [0.7854,0,-0.5236,0]$ and $\bm{u}_{\textrm{stat}} = [26.1239,9.4757]$, which is the initial point considered in Ref.~\cite{Luus2019}.
This approximation is denoted as \taylor.
We obtain $n_{\textrm{sim}} = 100$ different simulations for the system for perturbations ($\delta = 0.1$) around the expansion point to construct \linfit{}.
Since constructing a linear-fit model is computationally inexpensive, no subsampling is performed, and all the training data is used.
For example, it takes $0.0068$ seconds to construct \linfit{} for nearly one million $\langle\bm{I},\dot{\bm{X}} \rangle$ samples.

By comparing the derivatives predicted by these two linear approximations against \actual~and with each other, we can validate the efficacy of using the linear-fit model.
The derivatives as predicted by \taylor{} and \linfit{} for one of the test simulations are shown in Fig.~\ref{fig:dx_lin_comp}.
From Eq.~(\ref{eq:two-link-robot}), we can see $\dot{\xi}_1$ and $\dot{\xi}_3$ are linear with respect to the states.
From Figs.~\ref{fig:dx_lin_1} and \ref{fig:dx_lin_3}, we can see that \taylor{} and \linfit{} can capture this relation accurately.
If \linfit{} can accurately predict some state derivatives, then we do not construct $\bm{e}$.
This insight can lower the time required to construct the DFSM model.
Correspondingly, there is some error with the predictions for $\dot{\xi}_2$ and $\dot{\xi}_4$ as a linear model cannot accurately capture the nonlinearities.
This error keeps increasing as we move away from the expansion point (as is expected for linear models), and $\bm{e}$  would be necessary if a more accurate response is desired.



\begin{figure}[t]
    \centering
    \includegraphics[width = 0.33\textwidth]{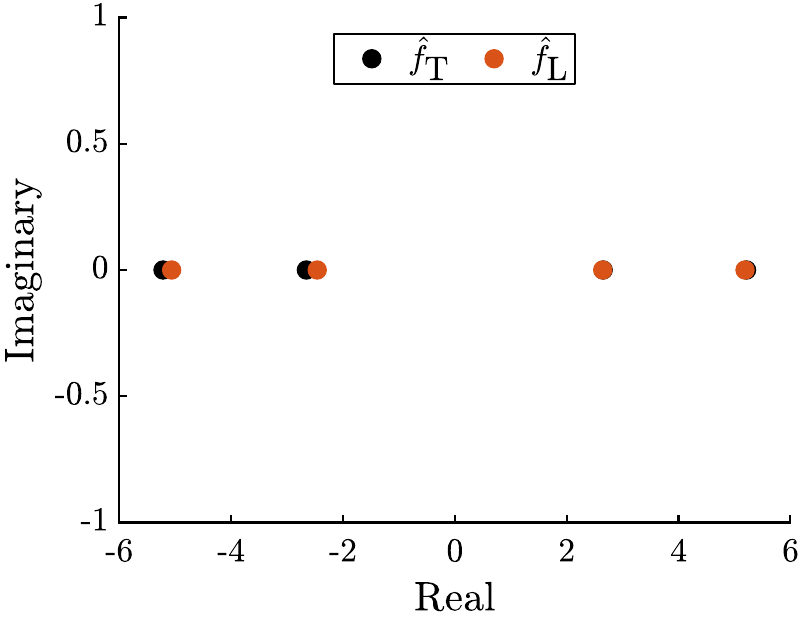}
    \caption{Comparisons of the eigenvalues (poles) of \linfit{} and \taylor{}.}
    \label{fig:poles-comparison}
\end{figure}

Looking at the eigenvalues of $\bm{A}_L$ (or the closely-related poles of the transfer function) is another way comparing \linfit{} to \taylor.
The poles for \taylor{} are at $[\pm 5.2160, \pm 2.6533]$.
The eigenvalues from \linfit{} are plotted in Fig.~\ref{fig:poles-comparison}.
By sampling around the expansion point, a good approximation of \taylor{} is constructed.
Considering the ease of constructing the \linfit{} model, this insight can be used to approximate the first-order Taylor series expansion.
However, care must be taken when approximating certain systems this way.
For example, a similar two-link-robot system, discussed in Section 12.4.2 of Ref.~\cite{Luus2019}, has poles at $\bm{s} = 0$.
For such a system, trying to match the poles of \linfit{} to \taylor{} can be numerical challenging.
\subsubsection{Validating the Multi-Fidelity Approach.}~To validate the multi-fidelity approach, we extend the set of $n_{\textrm{sim}} = 100$ simulations to 5 seconds to construct the DFSM over a much larger state-space range.
We use $80 \%$ of the simulations to train the model and use the rest for testing.
From the simulated data, $n_s = 500$ samples are extracted using the $k$-means approach outlined in Sec.~\ref{sec:k-means-sampling}. 
When constructing $\bm{e}(\cdot)$, we evaluate the error between \linfit{} and \actual.
Only for the state derivatives with high error magnitude $(>10^{-5})$ do we construct $e_i$ (i.e., only $\xi_2$ and $\xi_4$ here).

\begin{figure*}[ht]
\centering
\begin{subfigure}[t]{0.25\textwidth}
\centering
\noindent\includegraphics[scale=0.34]{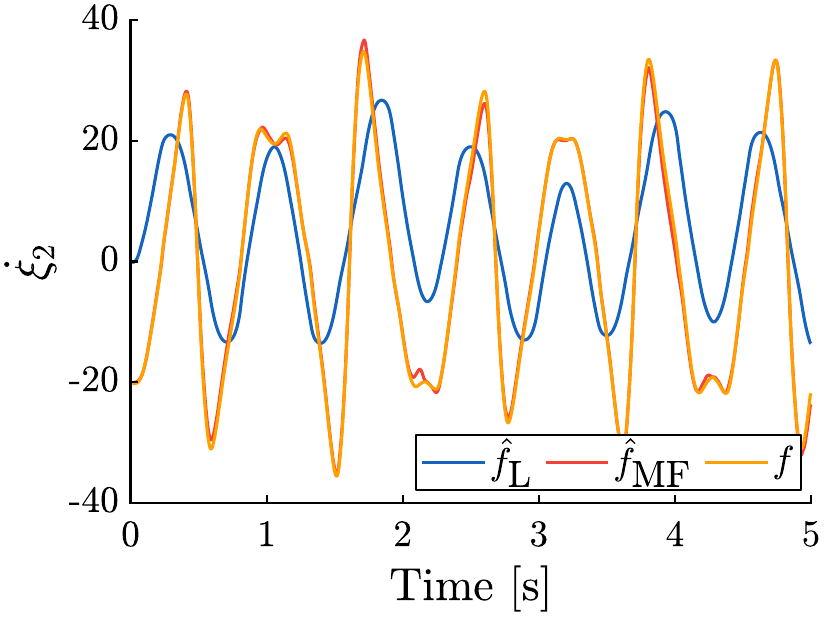}
\caption{$\dot{\xi}_2(t)$.}
\label{fig:dx_mf_2} 
\end{subfigure}%
\begin{subfigure}[t]{0.25\textwidth}
\centering
\noindent\includegraphics[scale=0.34]{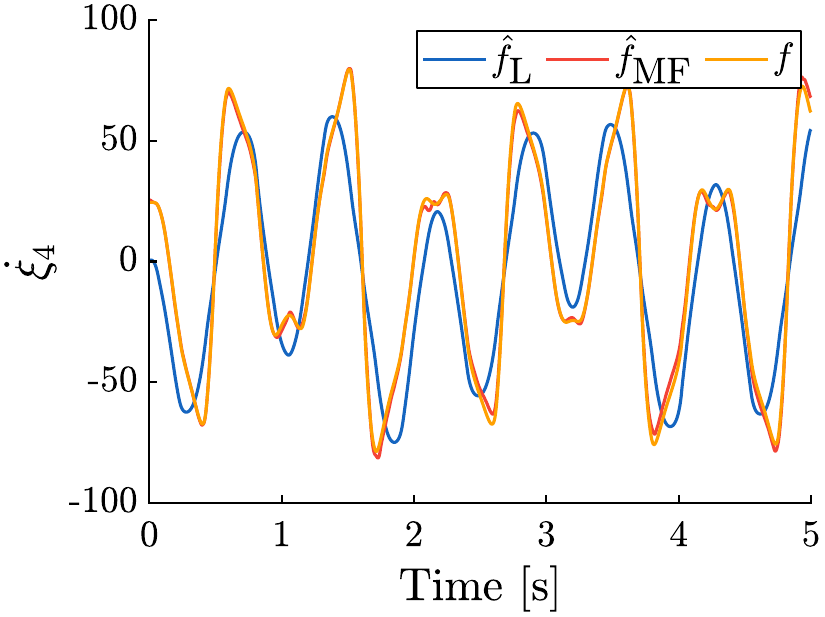}
\caption{$\dot{\xi}_4(t)$.}
\label{fig:dx_mf_4} 
\end{subfigure}%
\begin{subfigure}[t]{0.25\textwidth}
\centering
\noindent\includegraphics[scale=0.34]{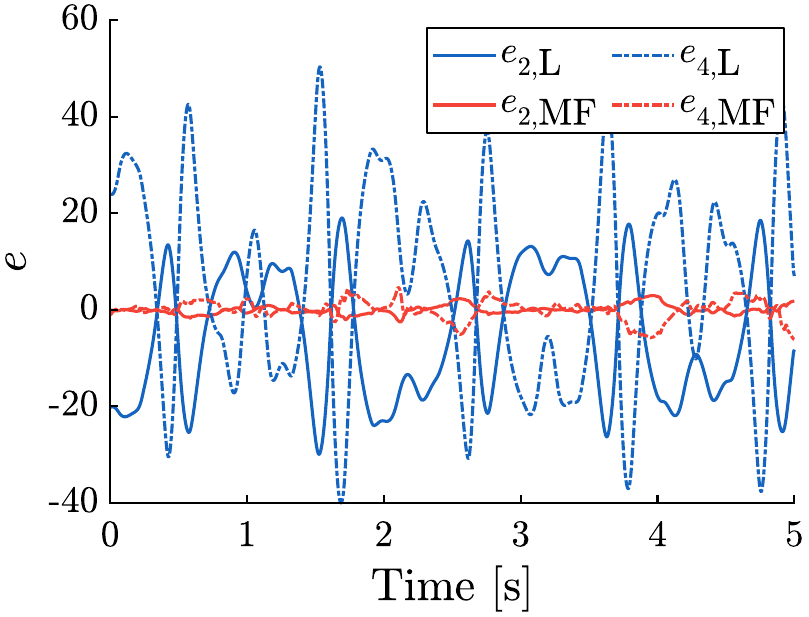}
\caption{${e}(t)$.}
\label{fig:dx_mf_e} 
\end{subfigure}%
\begin{subfigure}[t]{0.25\textwidth}
\centering
\noindent\includegraphics[scale=0.34]{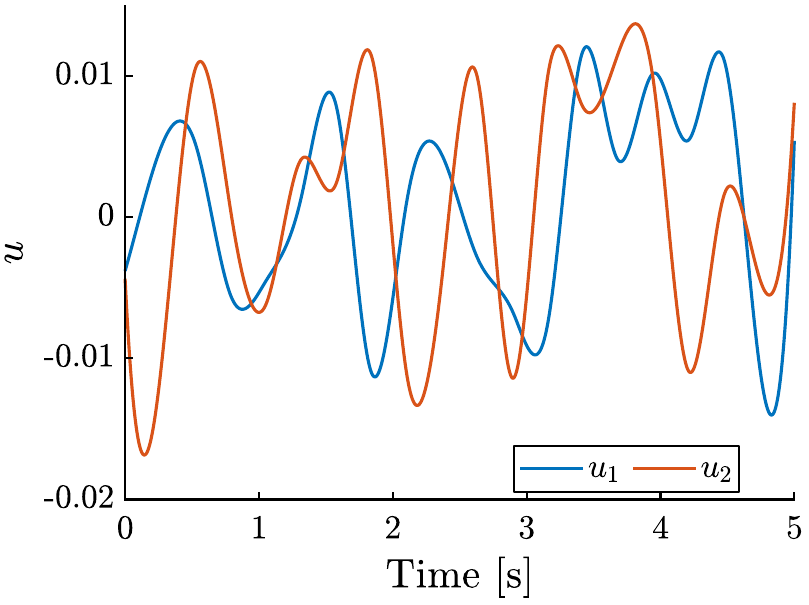}
\caption{$\bm{u}(t)$.}
\label{fig:u-input} 
\end{subfigure}%
\caption{Comparison of the state derivatives predicted by  \linfit~and \multifid~to \actual~for a test simulation using $\bm{u}(t)$.}
\label{fig:dx_mf_comp}
\end{figure*}
\begin{figure*}[ht]
\centering
\includegraphics[height=0.22in]{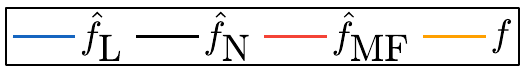}\\
\begin{subfigure}[t]{0.25\textwidth}
\centering
\noindent\includegraphics[scale=0.34]{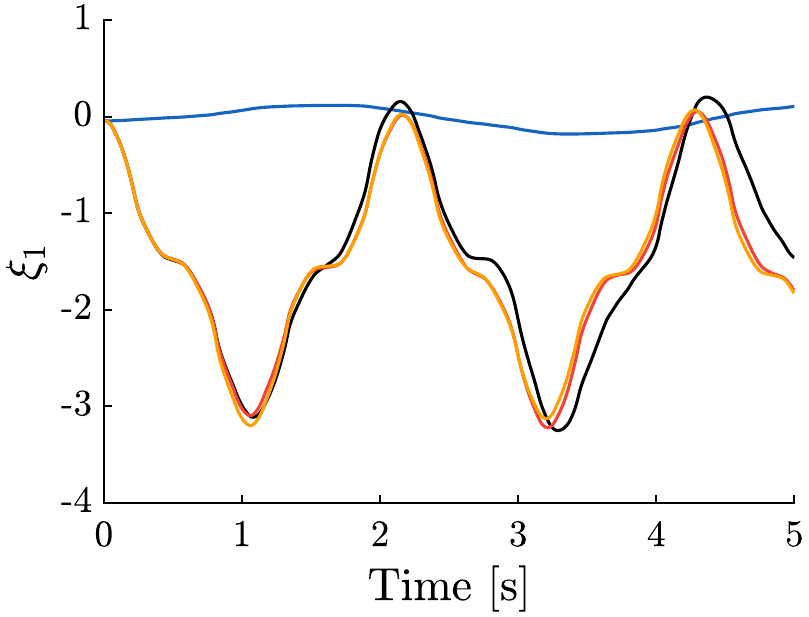}
\caption{$\xi_1(t)$.}
\label{fig:x_mf_1} 
\end{subfigure}%
\begin{subfigure}[t]{0.25\textwidth}
\centering
\noindent\includegraphics[scale=0.34]{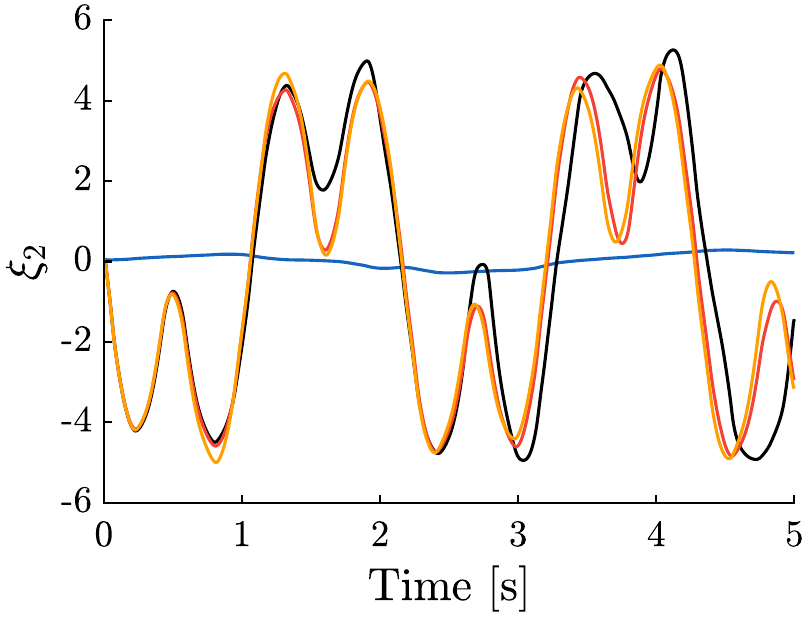}
\caption{$\xi_2(t)$.}
\label{fig:x_mf_2} 
\end{subfigure}%
\begin{subfigure}[t]{0.25\textwidth}
\centering
\noindent\includegraphics[scale=0.34]{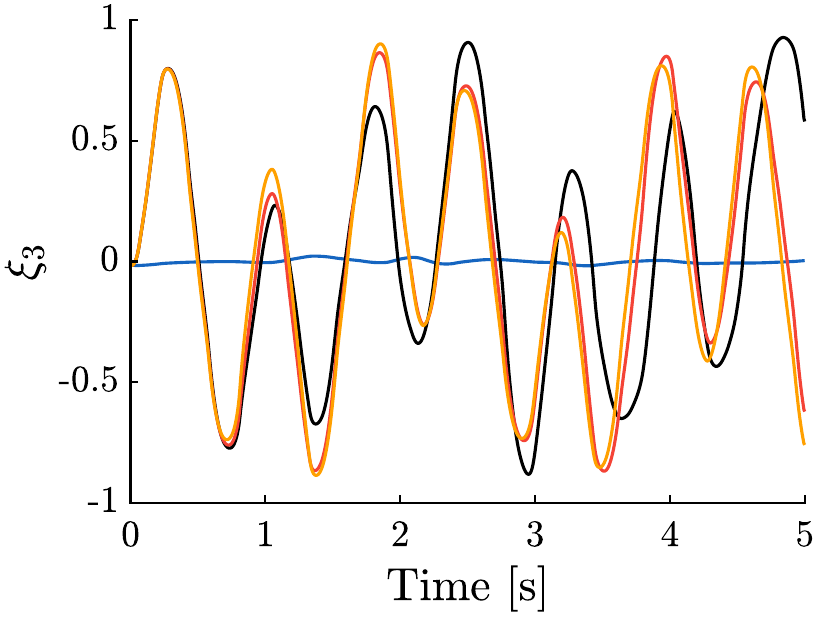}
\caption{$\xi_3(t)$.}
\label{fig:x_mf_3} 
\end{subfigure}%
\begin{subfigure}[t]{0.25\textwidth}
\centering
\noindent\includegraphics[scale=0.34]{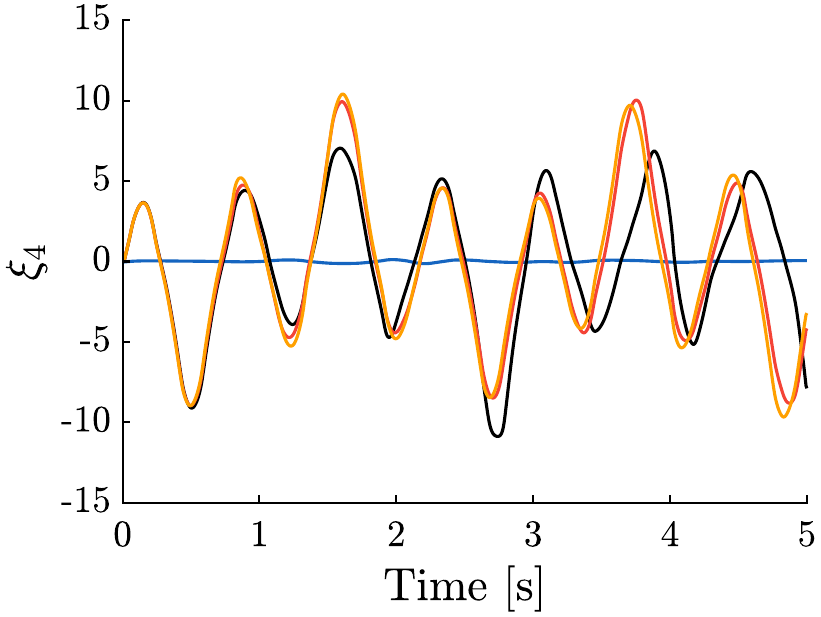}
\caption{$\xi_4(t)$.}
\label{fig:x_mf_4} 
\end{subfigure}%
\caption{Comparison of the states simulated using \linfit{} \multifid{} and \trad{} to \actual{} for a test simulation.}
\label{fig:x_mf_comp}
\end{figure*}
\begin{figure*}[ht]
\centering
\includegraphics[height=0.22in]{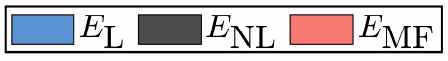}\\
\begin{subfigure}[t]{0.25\textwidth}
\centering
\noindent\includegraphics[scale=0.33]{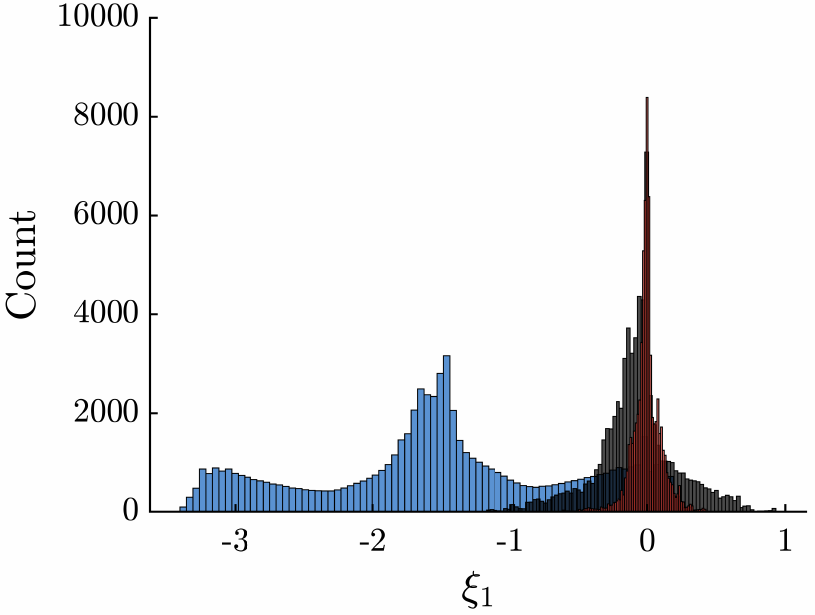}
\caption{Error in $\xi_1$.}
\label{fig:x_hist_1} 
\end{subfigure}%
\begin{subfigure}[t]{0.25\textwidth}
\centering
\noindent\includegraphics[scale=0.33]{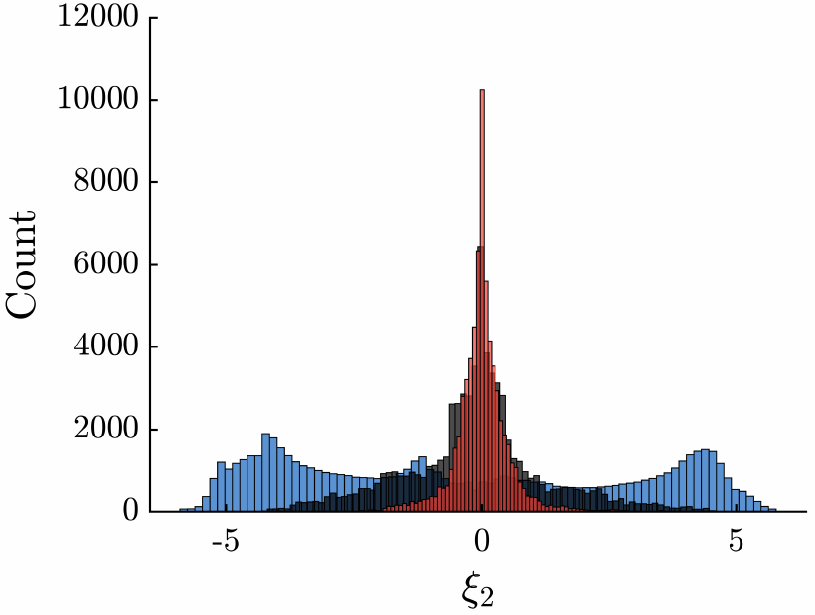}
\caption{Error in $\xi_2$.}
\label{fig:x_hist_2} 
\end{subfigure}%
\begin{subfigure}[t]{0.25\textwidth}
\centering
\noindent\includegraphics[scale=0.33]{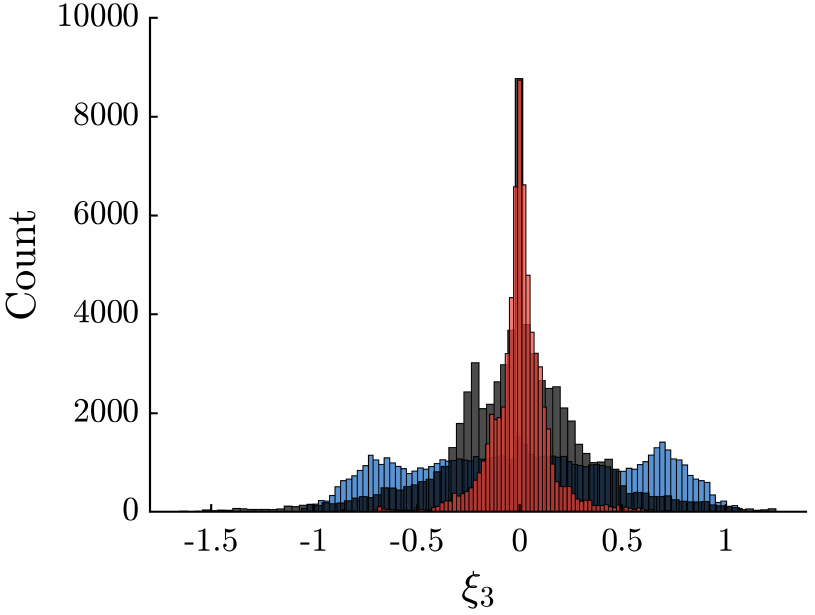}
\caption{Error in $\xi_3$.}
\label{fig:x_hist_3} 
\end{subfigure}%
\begin{subfigure}[t]{0.25\textwidth}
\centering
\noindent\includegraphics[scale=0.33]{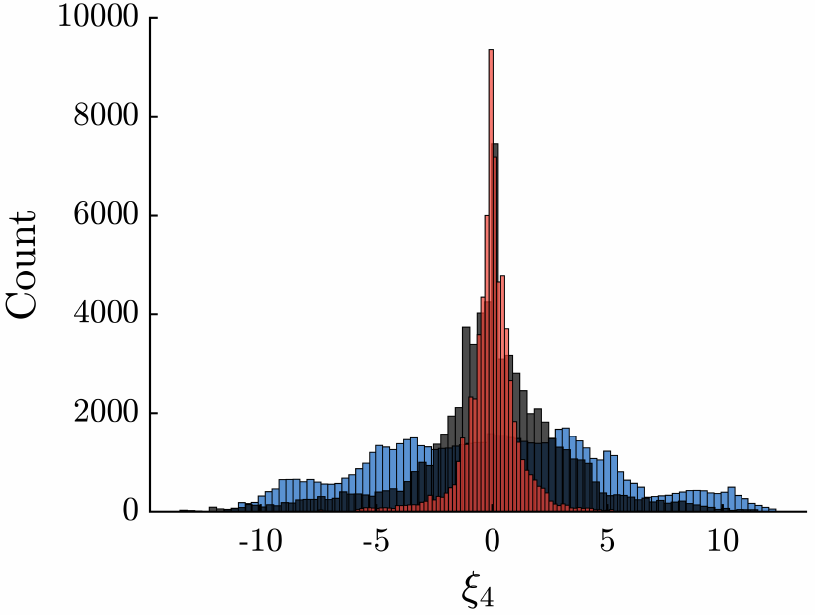}
\caption{Error in $\xi_4$.}
\label{fig:x_hist_4} 
\end{subfigure}%
\caption{Error histograms for the pointwise errors $E_{\textrm{NL}}$ between the states determined using a simulation with \actual{} versus with \trad{} and similar values $E_{\textrm{MF}}$ for \multifid{} for all test simulations.}
\label{fig:x_hist}
\end{figure*}




To understand the predictive capabilities of the DFSM and how the multi-fidelity approach compares to the traditional approaches, we simulate the system using \linfit{}, \multifid{}, and \trad{} and compare the response to the simulations using \actual.
To ensure a fair comparison, the same inputs and settings were used to construct $\bm{e}$ in \multifid~and the fully nonlinear model in \trad.
The derivatives predicted by both \linfit{} and \multifid{} (combination of \linfit{} and $\bm{e}$) are shown in Fig.~\ref{fig:dx_mf_comp} along with \actual.
With $\bm{e}$, the error between \multifid{} and \actual{} is significantly reduced, as shown in Figs.~\ref{fig:dx_mf_2} and \ref{fig:dx_mf_4}.
From the results shown in Fig.~\ref{fig:x_mf_comp}, we can see that \multifid{} can accurately capture the nonlinear response for this test simulation.
The states predicted by \trad{} are similar to \multifid, but the error increases as the simulation goes on.
The results predicted by using just \linfit{} are not accurate, showing the importance of a nonlinear model.
Time-domain simulations are an important test for a DFSM as the errors from previous time steps add up.
Even though the relation for $\dot{\xi}_1$ and $\dot{\xi}_3$ are known, the error from integrating the nonlinear derivatives $\dot{\xi}_2$ and $\dot{\xi}_4$ affects $\xi_1$ and $\xi_3$.

The results shown in Fig.~\ref{fig:x_mf_comp} are for a single test case.
To understand the model predictions for all test cases, we can look at the error histograms.
Accuracy can be inferred from the distribution of values around zero error.
Figure~\ref{fig:x_hist} shows histogram plots of the error in the states using different fidelity simulations for all twenty test trajectories (i.e., ones not used to train the model).
From the figures, we can see that \multifid{} is more accurate than \trad{} for the nonlinear derivatives for this test.
It is faster to construct \multifid{} compared to \trad{} since $\bm{e}$ need not be constructed for $\dot{\xi}_1$ and $\dot{\xi}_3$.
For different values of $n_s$, \multifid{} is on average $2$ times faster to construct than \trad.

As is the case for any metamodeling scheme, broad generalizations cannot be made from a single study.
For a different system, the trends seen in this example may not apply.
However, the multi-fidelity approach could be used to understand the trade-offs associated with approximating a complex (black box) function using a combination of simpler ones.
In cases where the multi-fidelity approach is less accurate than a fully nonlinear surrogate model, a linear-fit model can be used when necessary, and a nonlinear model can be constructed for more complex functions.
These studies also show that the method of extracting the derivative information and sampling the data from simulated results can be used to construct the DFSM using traditional approaches.
\xsection{FOWT DFSM Case Study}\label{sec:FOWT-application}

This section applies the multi-fidelity DFSM approach to the primary case study on the optimal control design for a FOWT.

\subsection{Wind Turbine Controls Overview}
In wind turbine controls, the generator torque and the blade pitching angles are the main control variables considered.
The value of these variables depends heavily on the wind speed input $w$ \cite{Pao2009, Abbas2021}.
The traditional operating region for wind turbines is between $w \in [3,25]$ [m/s], which is divided into multiple subregions based on the wind speed value.
There are three main regions of interest, the below-rated, transition, and rated regions.
More details regarding control design for wind turbines can be found in Refs.~\cite{Pao2009}.
We use model here a model of the IEA-15 MW FOWT~\cite{IEA15MW, Allen2020, Gaertner2020}.

Load cases in the transition/near-rated regions are more challenging for wind turbine controls as the largest tower base and rotor thrust loads are seen in this region.
Additionally, the power generated from this region is weighted higher in the levelized cost of energy~(LCOE) calculation~\cite{Sundarrajan2021}.
For these reasons, load cases from this region are one of the main design drivers, as the optimal design must have a suitable trade-off between load management and power generation.
The DFSM must intrinsically capture these trade-offs for it to be effectively used in optimal control and CCD studies.
Consideration of the platform pitch $\Theta_p$ is also important for FOWT operation. 
Wind inputs with three different average wind speed values are considered with $w_{\textrm{avg},1} = 6$ [m/s] (below-rated region), $w_{\textrm{avg},2} = 12$ [m/s] (near-rated/transition region), and  $w_{\textrm{avg},3} = 18$ [m/s] (rated region).

\subsection{Problem Formulation}

For this study, we consider a model with the platform pitch $\Theta_p$ and generator speed $\omega_g$ degrees of freedom enabled, so the state variables are $\Theta_p$ and $\omega_g$ as well as their first-order time derivatives.
The control inputs considered are the generator torque $\tau_g$ and the blade pitch $\beta$. A third input is the wind speed $w$. 
In addition to the states and controls, key outputs like the tower-base fore aft shear force $T_F$ and moment $T_M$ are also considered. Summarizing:
\begin{subequations}
\begin{align}
    \label{eq:FOWT-XU}
    \bm{\xi} &= [\Theta_p,\omega_g,\dot{\Theta}_p,\dot{\omega}_g]^T \\
    \bm{u} &= [\tau_g,\beta]^T \\
    \bm{y} & = [T_F,T_M]^T
\end{align}
\end{subequations}

\noindent The state derivatives are then:
\begin{align}
    \label{eq:FOWT-xdot}
    \dot{\bm{\xi}} &= [\dot{\Theta}_p,\dot{\omega}_g,\ddot{\Theta}_p,\ddot{\omega}_g]^T
\end{align}

\noindent The first two derivatives, $[\dot{\Theta}_p,\dot{\omega}_g]$, are the third and fourth states, so the linear-fit model would be sufficient to predict their values.

We use the WEIS toolbox available in Ref.~\cite{WEIS} to obtain $n_{\textrm{sim}} = 10$ simulations for DLC 1.1 for all three cases.
Then, the procedure outlined in Sec.~\ref{sec:DFSM} is used to construct individual DFSMs for each case with $n_s = 200$ samples used to construct $\bm{e}$ for both $\hat{\bm{f}}$ and $\hat{\bm{g}}$.
We use an LPV formulation for the \linfit{}, based on the wind speed $w$ according to Eq.~(\ref{eq:LPV-formulation}).
We use $80 \%$ of the simulations to construct the DFSM, and the remainder is used for testing.
We validate all three DFSMs using state simulations.

The main objective is to maximize the power generated.
Power is calculated as $p_g = \eta\tau_g\omega_g$, where $\eta = 0.99$ is the generator efficiency.
As explained in Refs.~\cite{Betts2010,Herber2014a,Sundarrajan2021}, since a control variable $\tau_g$ is linear in the objective function, a quadratic penalty must be added to prevent bang-bang behavior.
In addition to this, we also add a penalty on the blade pitch to explore its limited actuation in the transition region.
The single objective can be formulated as:
\begin{align}
    \label{eq:FOWT-objective}
    J = \underset{\bm{u},\bm{\xi}}{\text{minimize:}} \quad \int_{t_0}^{t_f} \left[  (-p_g) + w_1\tau_g^2 + w_2\beta^2 \right] \mathrm{d}t
\end{align}

\noindent Simple linear bound constraints are included to limit the values of key signals to within their rated values:
\begin{subequations}
\begin{align}
    \label{FOWT-constraints}
    & [1.67,0] \leq \bm{u} \leq [19.9,22.6] \quad [\textrm{MNm,deg}]\\
    & [0,1.9] \leq \bm{\xi} \leq [6,7.2]  \quad [\textrm{deg,rpm}] \\
    & [0,0] \leq \bm{y} \leq [5,4] \quad [\textrm{MN,MNm}]
\end{align}
\end{subequations}

The optimal control problem is formulated and solved using \texttt{DTQP}, an open source \texttt{MATLAB} toolbox that uses the DT method to discretize the problem~\cite{Herber2020d, dtqp}.
Then, \texttt{fmincon} with an interior-point method is used to solve the discretized nonlinear program.
A total of $n_t = 1000$ points are used to discretize the problem.
We set an optimality tolerance of $10^{-7}$ and use the central-finite difference method to evaluate the derivatives of the objective and constraints.

\subsection{Results}
\begin{figure*}
\centering
\begin{subfigure}[b]{0.32\textwidth}
    \centering
    \includegraphics[scale=0.41]{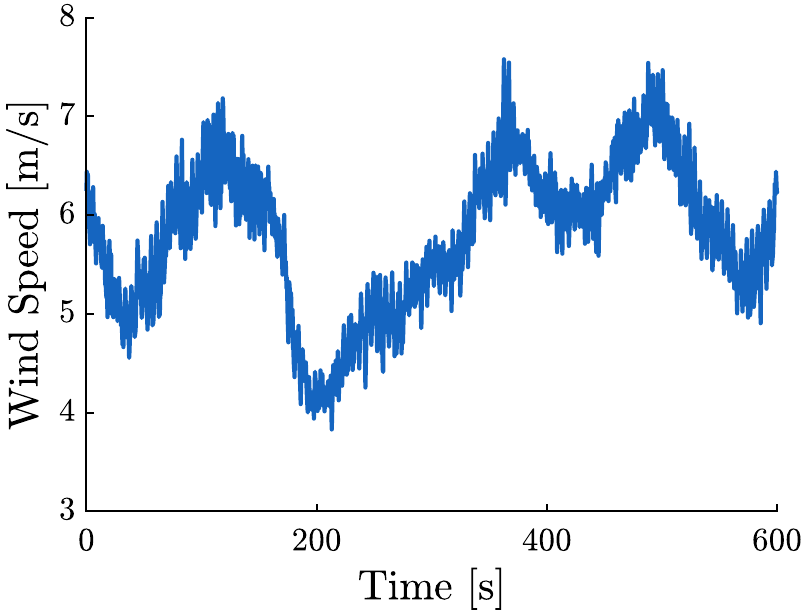}
    \caption{Wind speed trajectory with $w_{\textrm{avg},1}$.}
    \label{fig:wind_speed_1}
\end{subfigure}%
\hspace{0.0001\textwidth}%
\begin{subfigure}[b]{0.32\textwidth}
    \centering
    \includegraphics[scale=0.41]{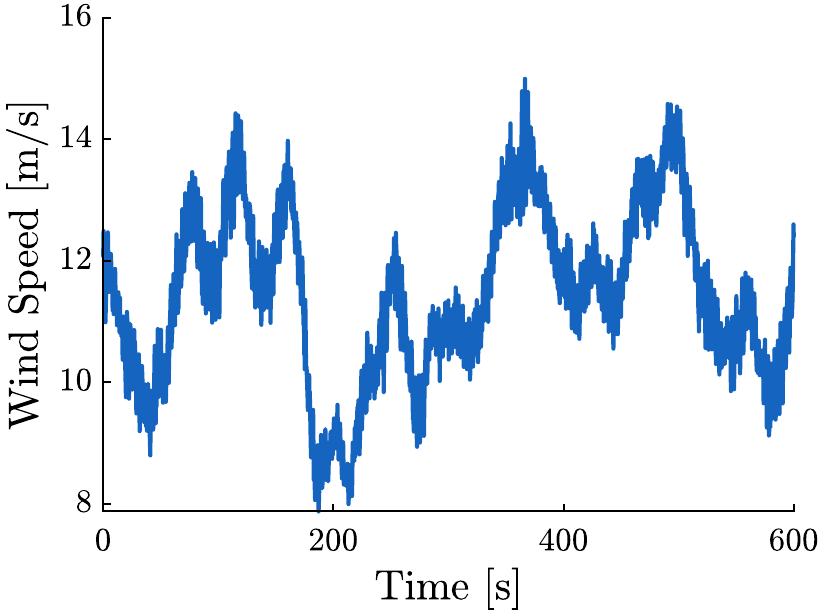}
    \caption{Wind speed trajectory with $w_{\textrm{avg},2}$.}
    \label{fig:wind_speed_2}
\end{subfigure}%
\hspace{0.0001\textwidth}%
\begin{subfigure}[b]{0.32\textwidth}
    \centering
    \includegraphics[scale=0.41]{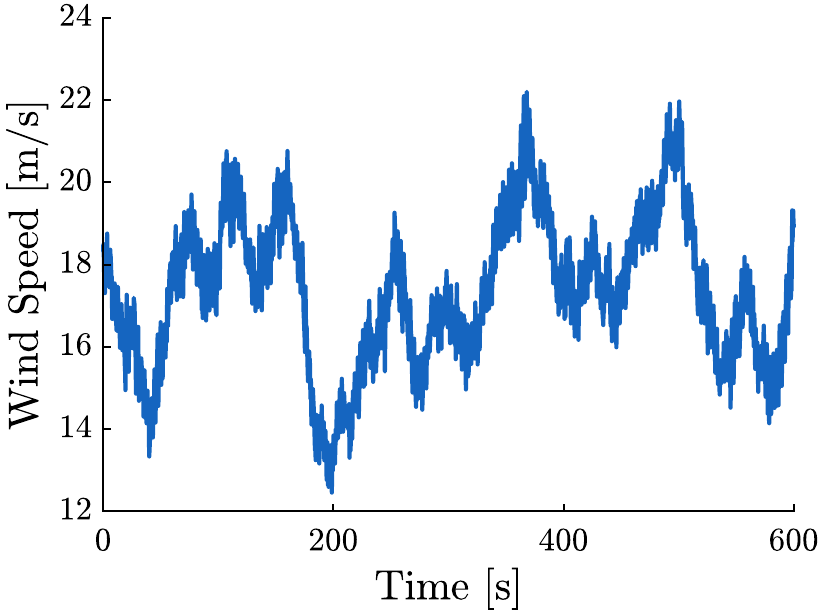}
    \caption{Wind speed trajectory with $w_{\textrm{avg},3}$.}
    \label{fig:wind_speed_3}
\end{subfigure}\\
\includegraphics[height=0.22in]{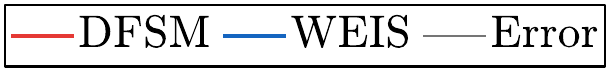}\\
\begin{subfigure}[b]{0.32\textwidth}
    \centering
    \includegraphics[scale=0.41]{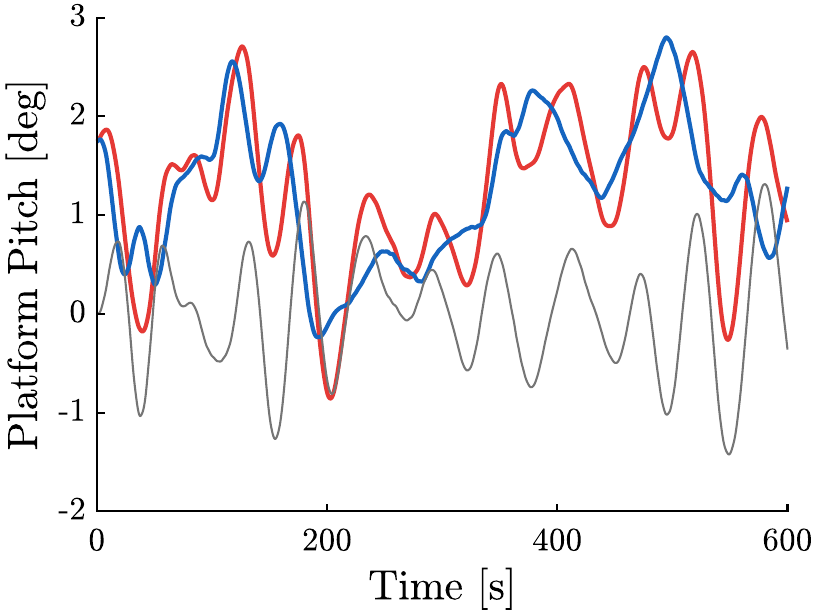}
    \caption{Platform pitch comparison for $w_{\textrm{avg},1}$.}
    \label{fig:ptfm_pitch_1}
\end{subfigure}%
\hspace{0.0001\textwidth}%
\begin{subfigure}[b]{0.32\textwidth}
    \centering
    \includegraphics[scale=0.41]{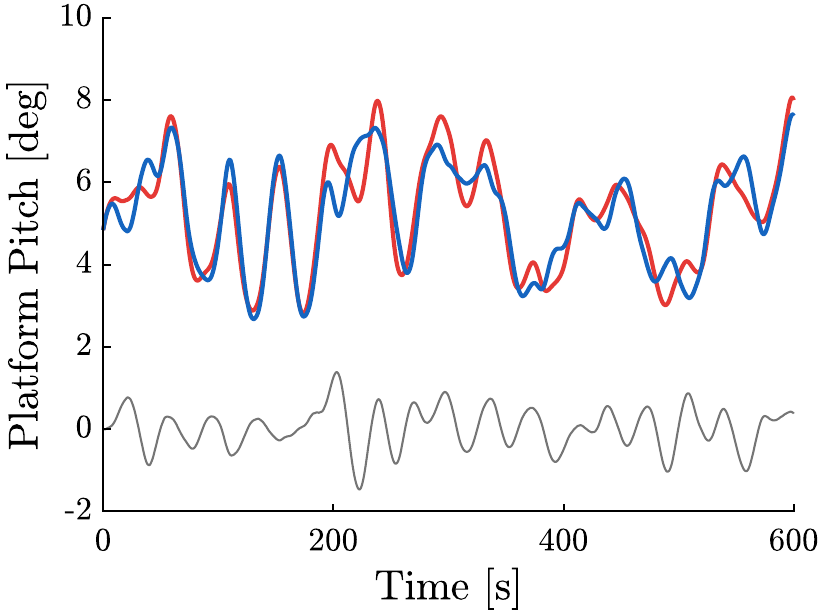}
    \caption{Platform pitch comparison for $w_{\textrm{avg},2}$.}
    \label{fig:ptfm_pitch_2}
\end{subfigure}%
\hspace{0.0001\textwidth}%
\begin{subfigure}[b]{0.32\textwidth}
    \centering
    \includegraphics[scale=0.41]{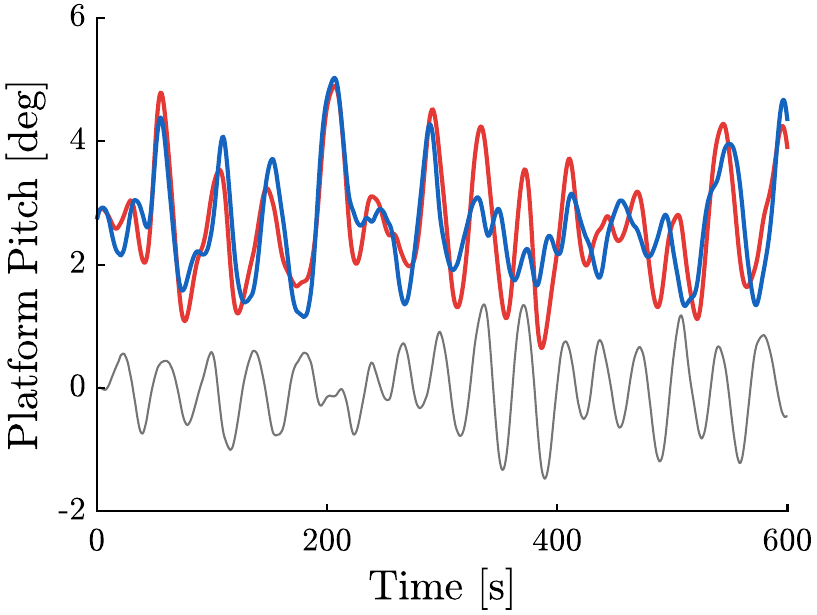}
    \caption{Platform pitch comparison for $w_{\textrm{avg},3}$.}
    \label{fig:ptfm_pitch_3}
\end{subfigure}\\
\begin{subfigure}[b]{0.32\textwidth}
    \centering
    \includegraphics[scale=0.41]{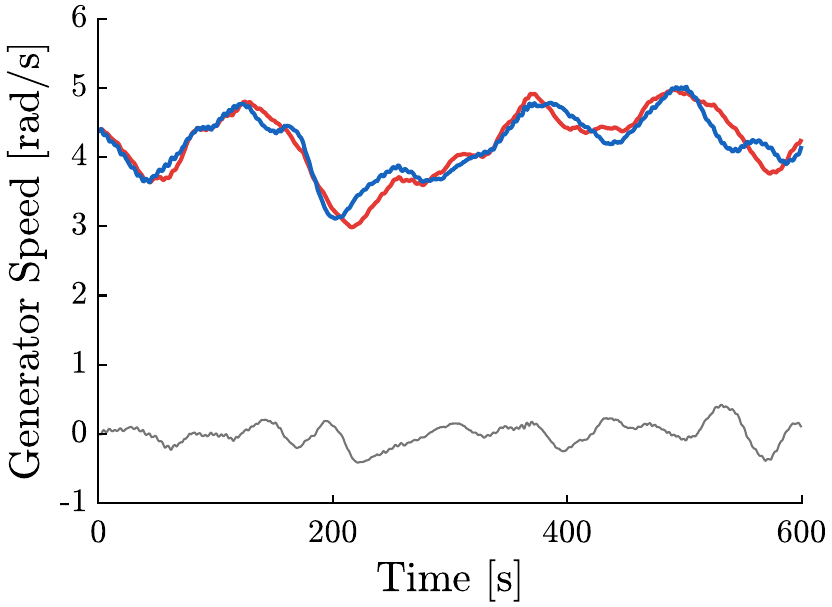}
    \caption{Generator speed comparison for $w_{\textrm{avg},1}$.}
    \label{fig:gen_speed_1}
\end{subfigure}%
\hspace{0.0001\textwidth}%
\begin{subfigure}[b]{0.32\textwidth}
    \centering
    \includegraphics[scale=0.41]{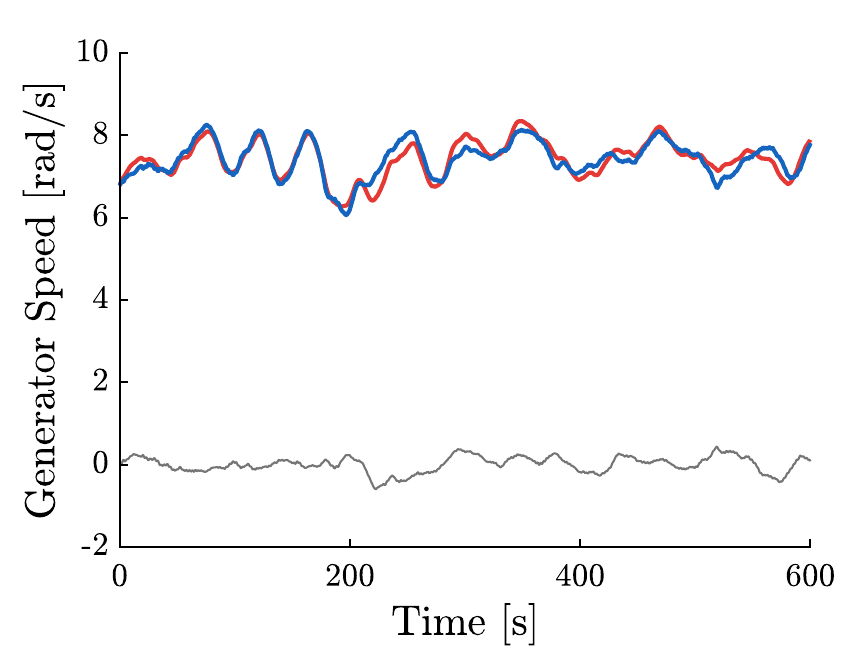}
    \caption{Generator speed comparison for $w_{\textrm{avg},2}$.}
    \label{fig:gen_speed_2}
\end{subfigure}%
\hspace{0.0001\textwidth}%
\begin{subfigure}[b]{0.32\textwidth}
    \centering
    \includegraphics[scale=0.41]{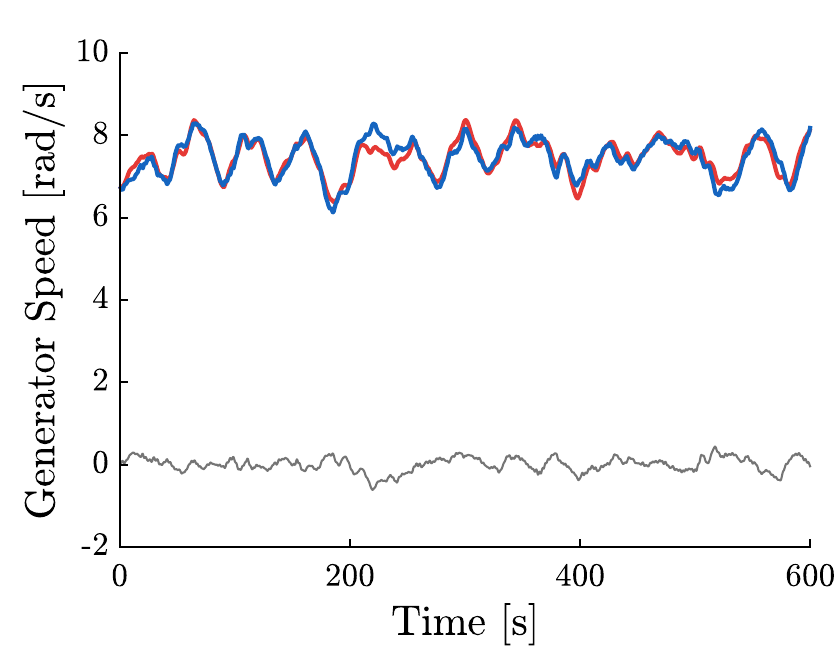}
    \caption{Generator speed comparison for $w_{\textrm{avg},3}$.}
    \label{fig:gen_speed_3}
\end{subfigure}\\
\caption{Validation results for \multifid{} for three different $w_{\textrm{avg}}$ values.}
\label{fig:FOWT_validation}
\end{figure*}
\begin{figure*}
\centering
\includegraphics[height=0.22in]{input/plots/fig_FOWT_validation/legend_common.pdf}\\
    \begin{subfigure}[b]{0.32\textwidth}
    \centering
    \includegraphics[scale=0.41]{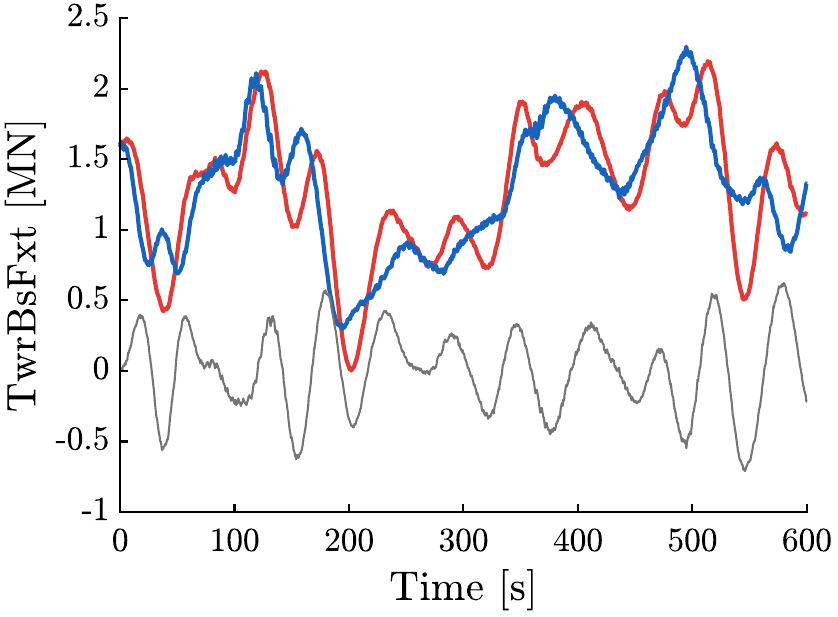}
    \caption{Tower base fore-aft shear force for $w_{\textrm{avg},1}$.}
    \label{fig:twrbs_fxt_1}
\end{subfigure}%
\hspace{0.0001\textwidth}%
\begin{subfigure}[b]{0.32\textwidth}
    \centering
    \includegraphics[scale=0.41]{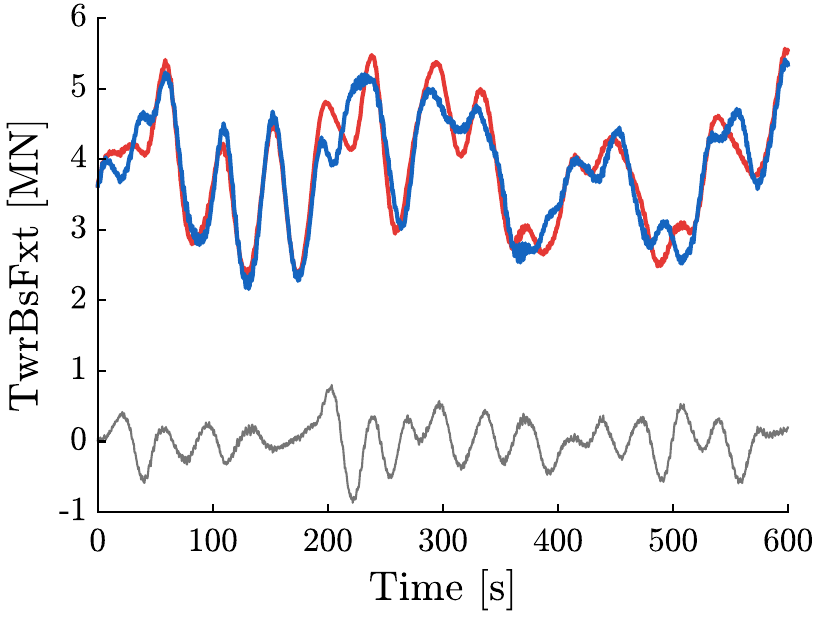}
    \caption{Tower base fore-aft shear force for $w_{\textrm{avg},2}$.}
    \label{fig:fig:twrbs_fxt_2}
\end{subfigure}%
\hspace{0.0001\textwidth}%
\begin{subfigure}[b]{0.32\textwidth}
    \centering
    \includegraphics[scale=0.41]{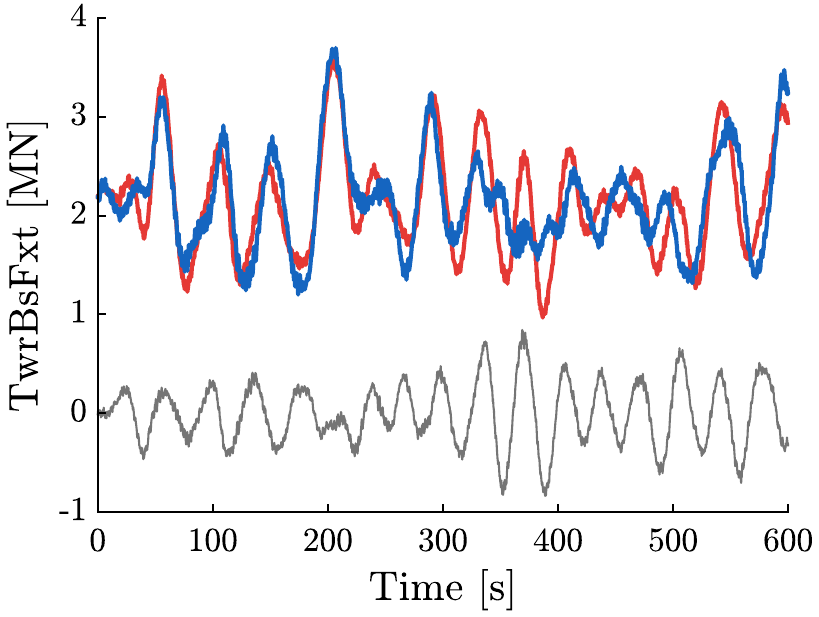}
    \caption{Tower base fore-aft shear force for $w_{\textrm{avg},3}$.}
    \label{fig:fig:twrbs_fxt_3}
\end{subfigure}\\
\begin{subfigure}[b]{0.32\textwidth}
    \centering
    \includegraphics[scale=0.41]{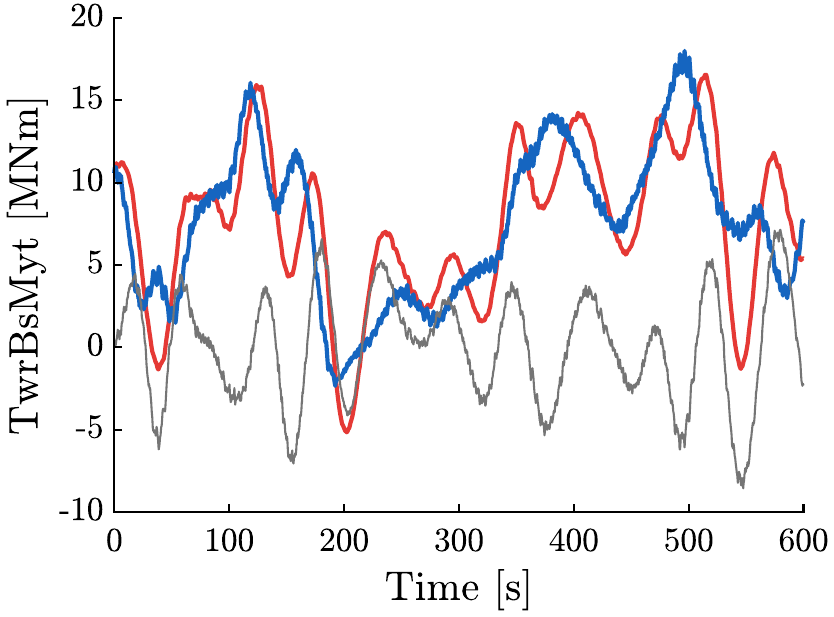}
    \caption{Tower base fore-aft shear moment for $w_{\textrm{avg},1}$.}
    \label{fig:fig:twrbs_mxt_1}
\end{subfigure}%
\hspace{0.0001\textwidth}%
\begin{subfigure}[b]{0.32\textwidth}
    \centering
    \includegraphics[scale=0.41]{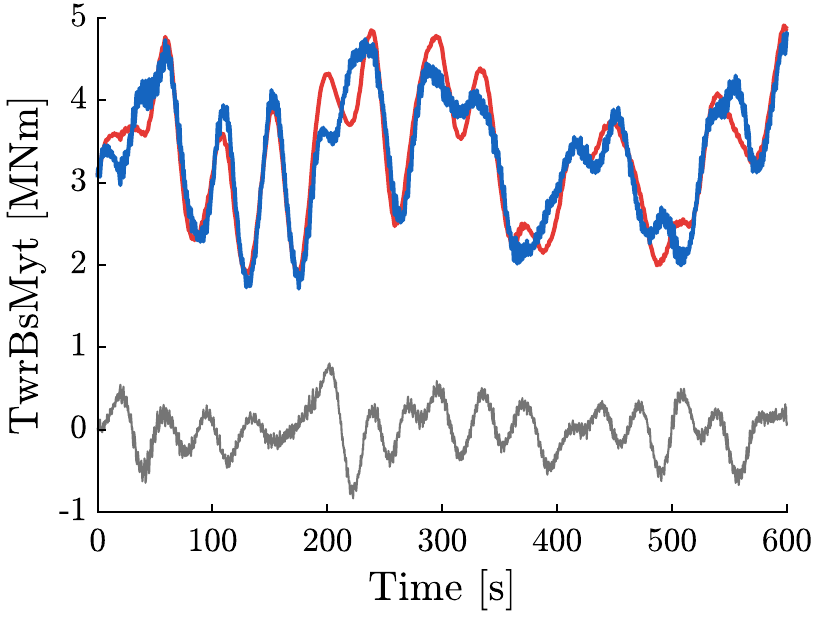}
    \caption{Tower base fore-aft shear moment for $w_{\textrm{avg},2}$.}
    \label{fig:fig:twrbs_mxt_2}
\end{subfigure}%
\hspace{0.0001\textwidth}%
\begin{subfigure}[b]{0.32\textwidth}
    \centering
    \includegraphics[scale=0.41]{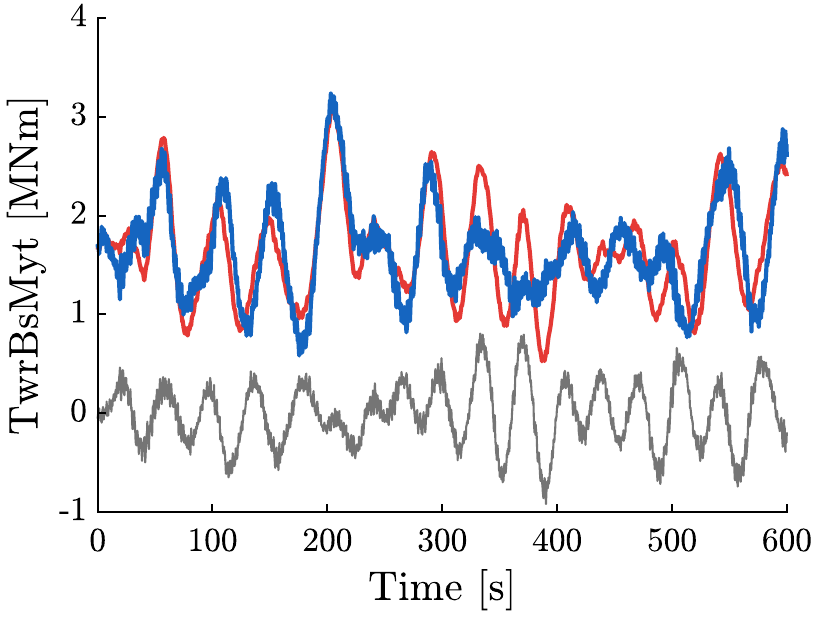}
    \caption{Tower base fore-aft moment for $w_{\textrm{avg},3}$.}
    \label{fig:fig:twrbs_mxt_3}
\end{subfigure}
\caption{Validation results for $\hat{\bm{g}}$ for three different $w_{\textrm{avg}}$ values.}
\label{fig:FOWT_validation_g}
\end{figure*}

The validation results for one of the test simulations for all three cases are presented in Fig.~\ref{fig:FOWT_validation}.
The wind speed trajectories are shown in Figs.~\ref{fig:wind_speed_1}--\ref{fig:wind_speed_3}.
Simulating a single load case takes an average of $20$ minutes.
However, once available, it takes an average time of $1.09$ minutes to extract the samples and construct the DFSM.
In addition to this, it takes an average of $4$ minutes to simulate one of the test load cases.
For these results, the DFSM simulations are $5$ times faster than WEIS.
Looking at Figs.~\ref{fig:ptfm_pitch_1}--\ref{fig:gen_speed_3}, the DFSM can predict the generator speed with sufficient accuracy, but there are some more significant errors associated with the platform pitch with similar peak values.
The outputs similarly can be accurately captured as shown in Figs.~\ref{fig:twrbs_fxt_1}--\ref{fig:fig:twrbs_mxt_3}.

\begin{figure*}[ht]
\centering
\includegraphics[height=0.22in]{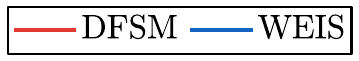}\\
    \begin{subfigure}[b]{0.32\textwidth}
    \centering
    \includegraphics[scale=0.41]{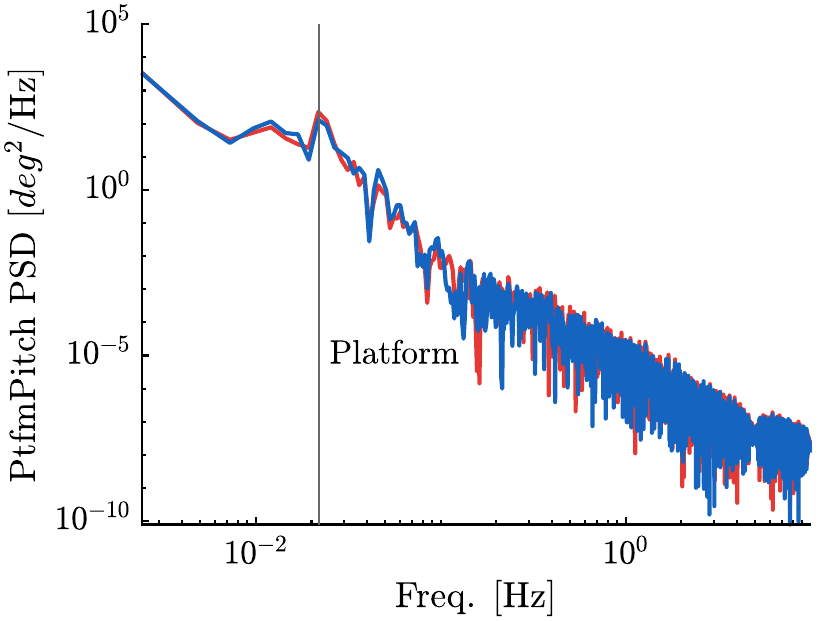}
    \caption{PSD of platform pitch.}
    \label{fig:ptfmpitch_FFT}
\end{subfigure}%
\hspace{0.0001\textwidth}%
\begin{subfigure}[b]{0.32\textwidth}
    \centering
    \includegraphics[scale=0.41]{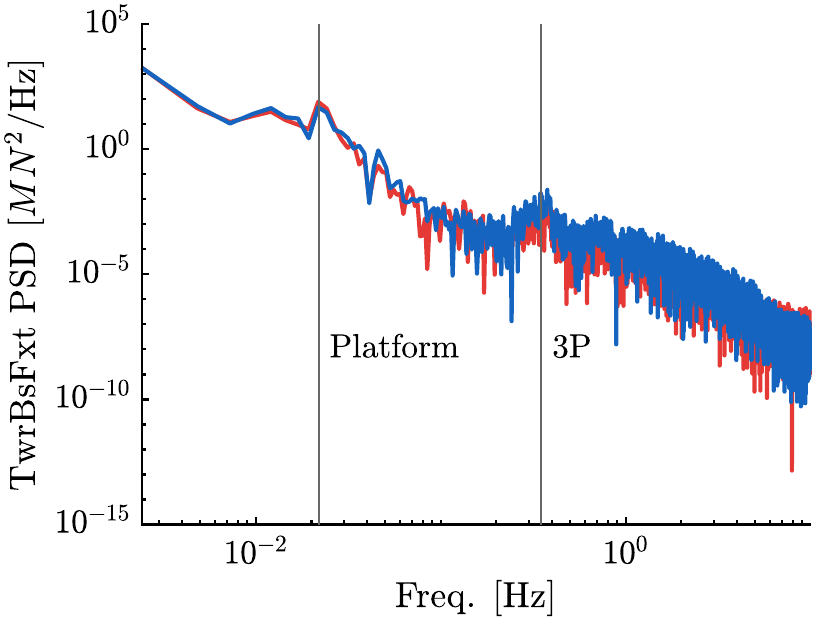}
    \caption{PSD of tower base fore-aft shear force.}
    \label{fig:twrbsfxt_FFT}
\end{subfigure}%
\hspace{0.0001\textwidth}%
\begin{subfigure}[b]{0.32\textwidth}
    \centering
    \includegraphics[scale=0.41]{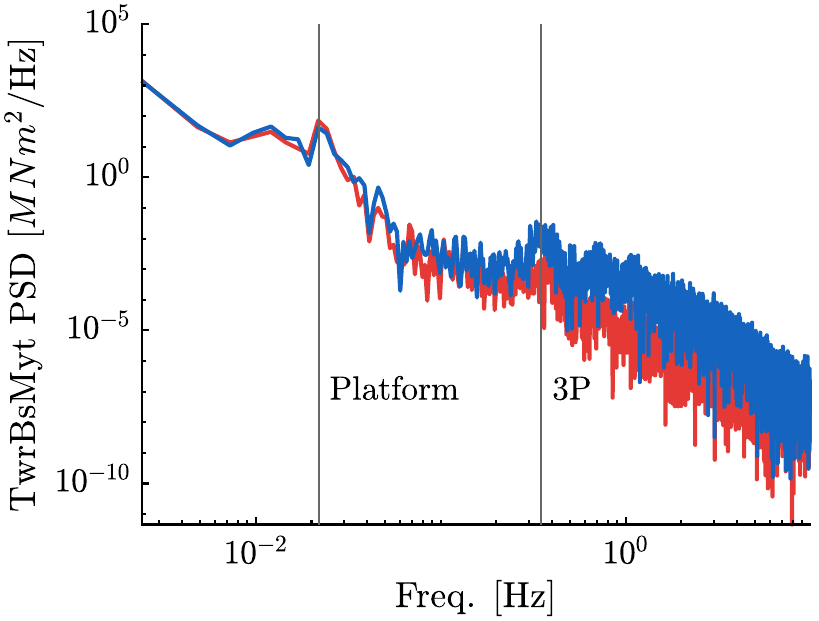}
    \caption{PSD of tower base fore-aft moment.}
    \label{fig:twrbsMxt_FFT}
\end{subfigure}
\caption{PSD of key signals from DFSM and WEIS in the transition region with $w_{\textrm{avg,2}}$.}
\label{fig:FOWT_validation_FFT}
\end{figure*}

In addition to the time-domain trends, the power spectral density (PSD) plots of key loads are studied~\cite{Solingen2014}.
A common control design goal is to minimize these loads in the transition region.
For the DFSM to be used in analysis and control design studies, it must capture the peaks in the PSD plot at key frequencies.
Figure~\ref{fig:FOWT_validation_FFT} shows the PSD plot of the platform pitch and the tower-base fore-aft shear force and moment from the DFSM and WEIS simulations for a test trajectory in the transition region.
The frequency range of $0-1$ Hz is crucial for wind turbine design, as most of the key system frequencies lie in this range~\cite{Gaertner2020, Allen2020}.
For a floating system, the platform pitching motion and the tower-base loads are affected by the natural frequency of the platform.
This frequency is plotted in Figs.~\ref{fig:ptfmpitch_FFT}--\ref{fig:twrbsMxt_FFT}.
Additionally, for the tower-base loads, another peak occurs at the 3P frequency (frequency at which the rotor blades pass the tower) as shown in Figs.~\ref{fig:twrbsfxt_FFT} and \ref{fig:twrbsMxt_FFT}.
Even though there is a slight difference between the power density value in this range, the peaks occur around the same key frequencies.
Since the goal is to use the DFSM rapidly in early-stage design studies, some accuracy loss might be expected as long as the DFSM can identify the right trade-offs.



\begin{figure*}[ht]
\centering
\includegraphics[height=0.22in]{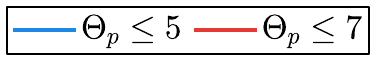}\\
\begin{subfigure}[t]{0.25\textwidth}
    \centering
    \includegraphics[scale=0.34]{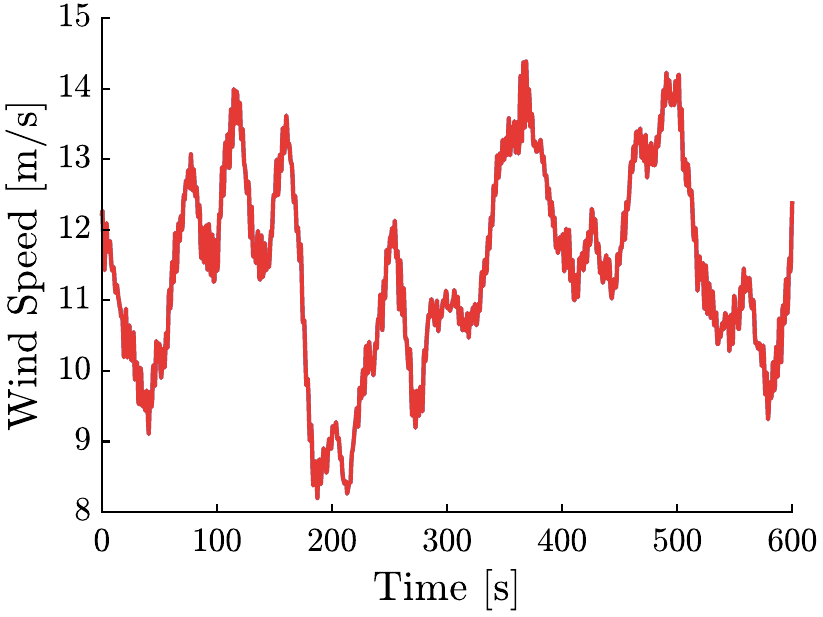}
    \caption{Wind speed ($w$).}
    \label{fig:wind_speed_transition}
\end{subfigure}%
\begin{subfigure}[t]{0.25\textwidth}
    \centering
    \includegraphics[scale=0.34]{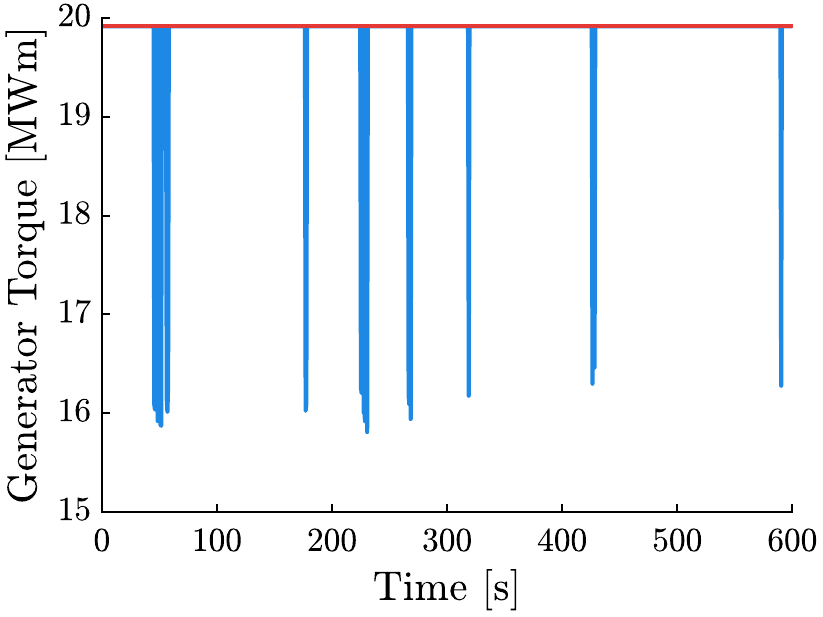}
    \caption{Generator torque ($\tau_g$).}
    \label{fig:gen_torque_transition}
\end{subfigure}%
\begin{subfigure}[t]{0.25\textwidth}
    \centering
    \includegraphics[scale=0.34]{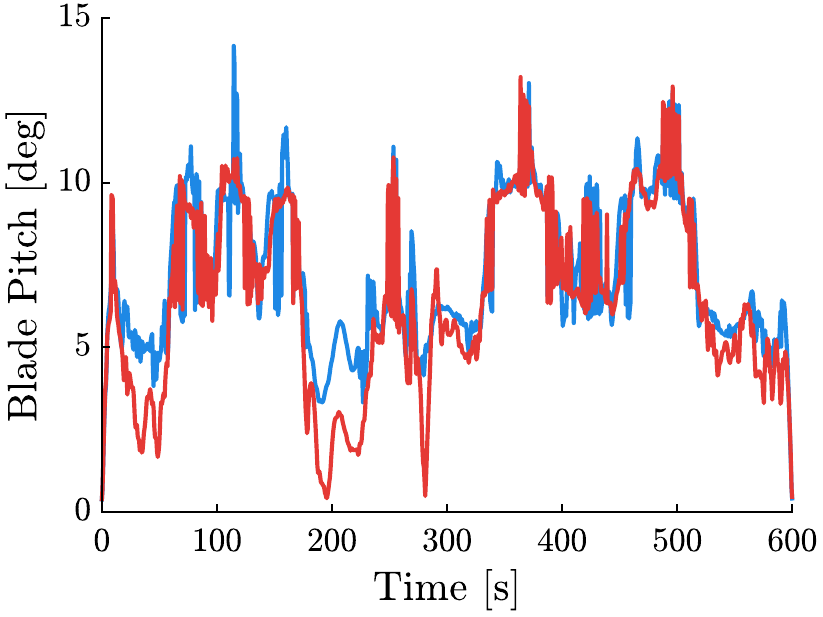}
    \caption{Blade pitch ($\beta$).}
    \label{fig:blade_pitch_transition}
\end{subfigure}%
\begin{subfigure}[t]{0.25\textwidth}
    \centering
    \includegraphics[scale=0.34]{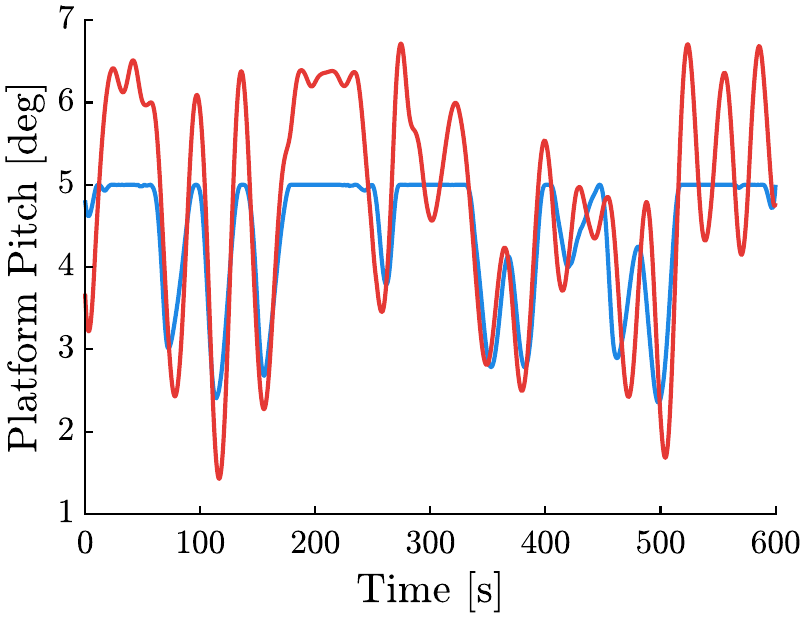}
    \caption{Platform pitch ($\Theta_p$).}
    \label{fig:ptfm_pitch_transition}
\end{subfigure}\\
\begin{subfigure}[t]{0.25\textwidth}
\centering
    \includegraphics[scale=0.34]{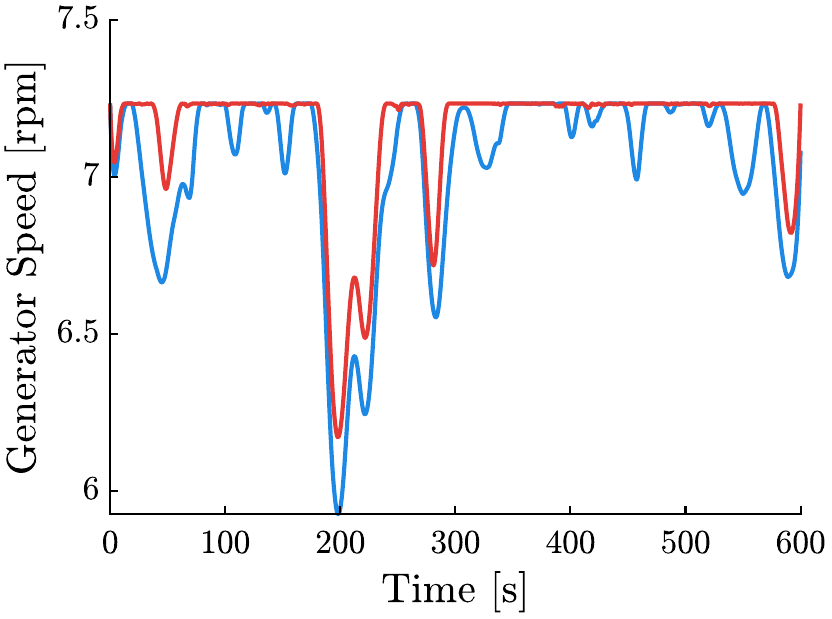}
    \caption{Generator speed ($\omega_g$).}
    \label{fig:gen_speed_transition}
\end{subfigure}%
\begin{subfigure}[t]{0.25\textwidth}
    \centering
    \includegraphics[scale=0.34]{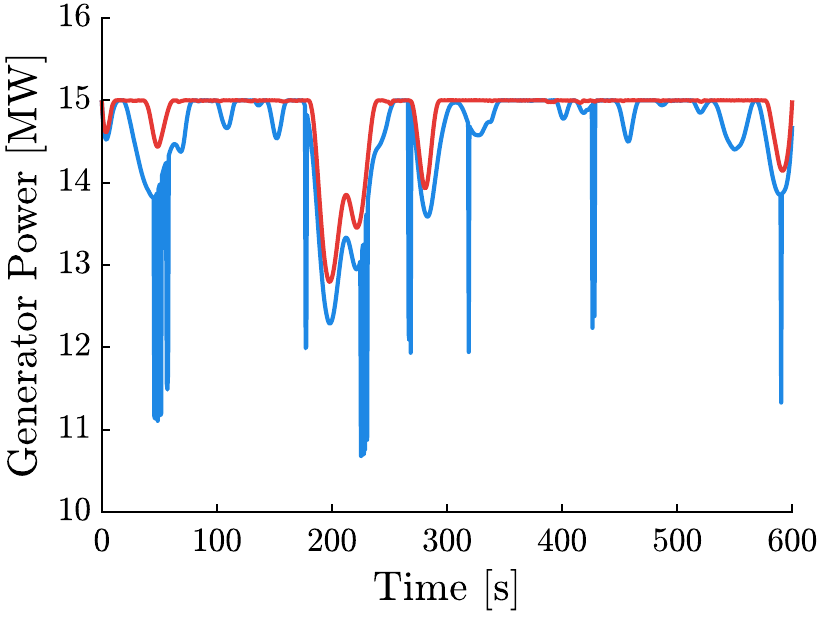}
    \caption{Generator power ($P = \eta \tau_g \omega_g$).}
    \label{fig:gen_power_transition}
\end{subfigure}%
\begin{subfigure}[t]{0.25\textwidth}
    \centering
    \includegraphics[scale=0.34]{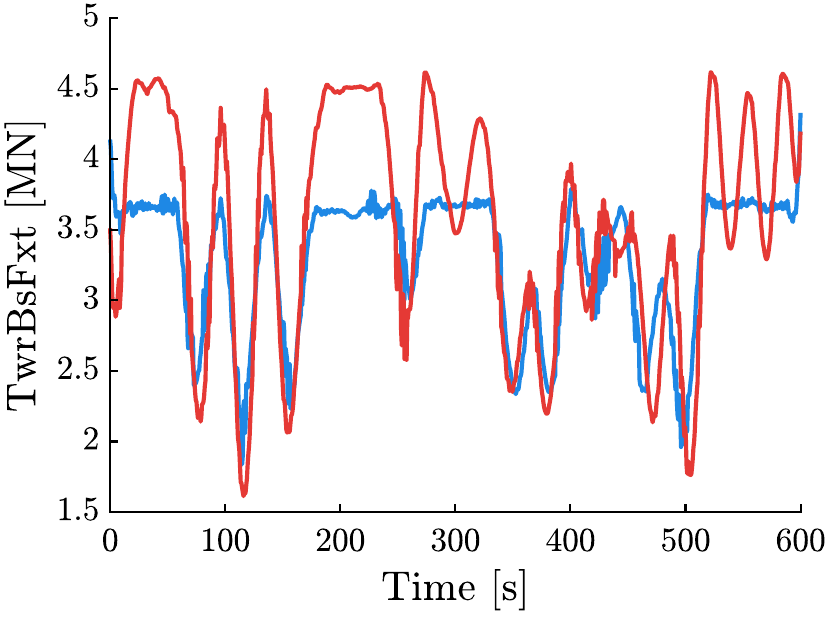}
    \caption{Shear force ($T_F$).}
    \label{fig:twrbsfxt_transition}
\end{subfigure}%
\begin{subfigure}[t]{0.25\textwidth}
    \centering
    \includegraphics[scale=0.34]{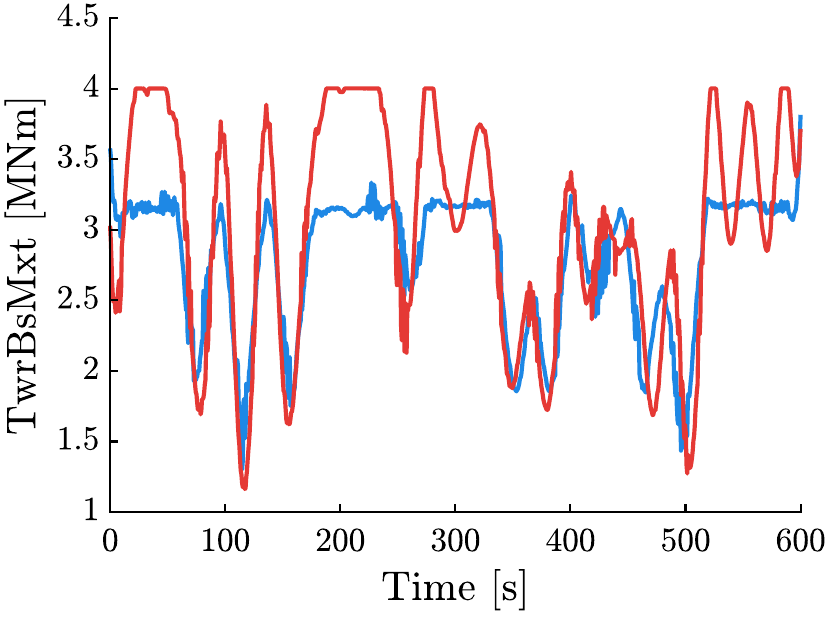}
    \caption{Shear moment ($T_M$).}
    \label{fig:twrbsmxt_transition}
\end{subfigure}%
\caption{Optimal control results using \multifid~for a single test load cases with $w_{\textrm{avg},2} = 12$ [m/s].}
\label{fig:oc_results_transition}
\end{figure*}

Optimal control results plotted in Fig.~\ref{fig:oc_results_transition} are for the load case in the transition region with $w_{\textrm{avg},2} = 12$ [m/s].
This problem was solved with $w_1 = 10^{-5}$ and $w_2 = 0.5$.
The optimal control problem was solved for two values of the $\Theta_{p,\max}$ as indicated for both the test load cases.
The results for one of the test cases are discussed.
The maximum platform pitch occurs around $w = 11$ [m/s].
To satisfy the constraint on $\Theta_{p,\max}$, blade pitch $\beta$ is active, which lowers the values of both $\Theta_p$, and $\omega_g$ as shown in Figs.~\ref{fig:ptfm_pitch_transition} and \ref{fig:gen_speed_transition}.
The generator speed is used to calculate the power, so the constraint on $\Theta_p$ affects the power generated.
For lower values of $\Theta_{p,\max}$, $\beta$ is more active, which lowers the power generated.
Similar results are observed for the second test case.
These results follow similar trends as the ones shown for similar studies in Ref.~\cite{Sundarrajan2021}. 

In addition to these results, optimal control problems were solved for wind inputs from the below-rated and rated regions.
These results are not directly shown, but the expected control trends were identified using the DFSM approach.
DLC 1.1 is used to test the power generation capability of wind turbines.
The results from this case study show that the DFSM, constructed using data from high-fidelity simulations, can be used to identify the right trade-offs.
Additional factors like pitch rate and tower top acceleration were not considered for this study, including hydrostatic and hydrodynamic constraints on the system.
These considerations will affect the optimal controls.

\xsection{Conclusion}\label{sec:conclusion}

In this article, we explored the use of a multi-fidelity derivative function surrogate modeling (DFSM) approach that can be used to approximate the dynamic model of nonlinear systems.
We proposed an approach to extract the state derivative information from system simulations by constructing polynomial approximations of the states and evaluating the derivatives of these approximations.
With the extracted state derivative information, the multi-fidelity DFSM consists of a least-squares linear-fit low-fidelity model and an additive nonlinear error corrective function to approximate the remaining error.

The response from the linear-fit model is compared against a first-order Taylor series expansion.
The results show that the linear-fit approximation can accurately find linear relations in the derivative function.
This feature helps lower the time required to construct the DFSM.
Then, we propose using simulations to validate the DFSM by comparing simulation results, which better accounts for the accumulation of state derivative function errors.
Finally, the DFSM approach was used to solve optimal control problems for floating offshore wind turbines~(FOWTs).
Inputs from several operating regions were tested.
Results show that using the DFSM results in well-known control trends and optimal trade-offs. 

Several improvements need to be made to this DFSM approach before it can be used as part of a control co-design (CCD) study.
The approach presented here must be extended to include plant variables as inputs to the DFSM.
To construct the DFSM for more inputs, scalable methods with respect to the number of inputs must be explored as alternatives to the radial basis functions~(RBF) used in this study.
An adaptive refinement method that can be used in conjunction with the optimal control study is needed to improve the accuracy of the DFSM.
Methods that can be used to obtain simulations that cover the entire state space of the given dynamic system would provide a better dataset for DFSM construction.
Different applications of this approach for designing hydrokinetic turbines, wave energy converters, and other related systems with expensive and potentially black box models can also be explored.

\begin{acknowledgment}
The information, data, or work presented herein was funded by the Advanced Research Projects Agency-Energy (ARPA-E), U.S. Department of Energy, under Award Number DE-AC36-08GO28308.
The authors would like to thank Dan Zalkind, Thanh Toan Tran, and Hannah Ross of NREL, and Saeed Azad of CSU for their feedback.
\end{acknowledgment}

\renewcommand{\refname}{REFERENCES}
\bibliographystyle{config/asmems4}
\begin{mySmall}
\bibliography{References}
\end{mySmall}



\end{document}